\documentclass[runningheads]{llncs}

\usepackage{eccv}

\usepackage{eccvabbrv}

\usepackage{graphicx}
\usepackage{wrapfig}
\usepackage{booktabs}

\usepackage[accsupp]{axessibility}  

\usepackage{hyperref}

\usepackage{orcidlink}

\usepackage{algorithm}
\usepackage{setspace}
\usepackage{algpseudocode}

\begin{document}


\title{Learning Manifolds in High-D Point Embedding for Anisotropic Surface Approximation from Unstructured Point Clouds} 

\titlerunning{HD-PEA}


\author{Hongbo Li\inst{1} \and
Haikuan Zhu\inst{1} \and
Xiaohu Guo\inst{2} \and Wenping Wang\inst{3} \and Jing Hua\inst{1} \and Zichun Zhong\inst{1} } 

\authorrunning{H. Li, et al.}

\institute{Department of Computer Science, Wayne State University, Detroit, USA\and
Department of Computer Science, Univeristy of Texas at Dallas, USA\and 
Department of Computer Science \& Engineering, Texas A\&M University, USA
\email{hm9026@wayne.edu}, \email{hkzhu@wayne.edu}, \email{xguo@utdallas.edu}, \email{wenping@tamu.edu}, \email{jinghua@wayne.edu}, \email{zichunzhong@wayne.edu}}

\maketitle

\begin{abstract}
Dense 3D sensors in various real-world fields produce point clouds that are geometrically redundant for real-time processing. In this paper, we propose an efficient and scalable learning-based anisotropic surface approximation framework, HD-PEA, that operates directly on unstructured point clouds, integrating anisotropic optimization into reconstruction to produce compact, geometry-aligned surface representations with higher fidelity, fewer elements, and improved numerical stability compared to isotropic and adaptive meshes. Firstly, we develop a novel learning-based high-dimensional (high-d) Euclidean point embedding method to map the input point clouds into a high-d manifold embedding space. For handling large-scale point clouds without retraining and fine-tuning, a patch-based meta-embedding scheme is designed during the inference stage. Then, we develop a new tangent subspace estimation for the high-d embedding manifold approximation and anisotropic manifold reconstruction in high-d space. The main contribution of this work is to propose a scalable deep learning framework and a variety of datasets for constructing a high-d Euclidean point embedding space aimed to 3D anisotropic surface mesh approximation and Riemannian curvature tensor estimation from point clouds. We extensively evaluate our method against state-of-the-art surface reconstruction approaches using several datasets, such as Thingi10K dataset, AIM@SHAPE and Stanford 3D Scanning Repository, ScanNet dataset, and further demonstrate its generalization and usability on diverse unseen shapes and applications from these datasets.
\end{abstract}


\section{Introduction}
\label{sec:intro}
In robotics, autonomous driving, and AR / VR applications, dense 3D sensors produce massive point clouds that are often too computationally expensive for real-time processing. Anisotropic surface meshes are therefore essential in geometric modeling, computer vision, and computer graphics, as they enable more accurate and efficient shape representations while significantly reducing computational cost. Unlike isotropic meshes, where the elements are as regular and uniform as possible, anisotropic surface mesh representation uses elongated triangles aligned with specific directions to optimize surface approximation, such as following the curvature tensor's eigenvalues and eigenvectors~\cite{Simpson:1994,Heckbert:1999}, leading to higher geometric accuracy with fewer elements and to improve downstream geometric processing efficiency. Different from existing anisotropic remeshing methods with the given input meshes~\cite{borouchaki1997delaunayI,du2005anisotropic,zhong2013,fu2014anisotropic,boissonnat2015anisotropic,Embedding2018,Li2024}, in this paper, we address a more challenging cross-modality problem in approximating the anisotropic surface mesh directly from unstructured 3D point clouds. The fundamental motivation of this work is to present a new manifold-guided surface reconstruction framework to the synergic integration of anisotropic meshing and surface reconstruction for effective 3D shape representation.

With the advancement of neural representations, learning implicit functions, such as signed distance functions (SDFs), from point clouds has become a prevalent approach to surface reconstruction. To achieve high reconstruction accuracy, they often rely on high-resolution Marching Cubes (MC) results~\cite{lorensen1998marching}, which generate densely sampled meshes with a large number of vertices and faces. 
Recent neural meshing methods, such as NMC~\cite{chen2021neural} and NDC~\cite{chen2022neural}, enhance Marching Cubes and Dual Contouring~\cite{ju2002dual} by learning vertex placements. VoroMesh~\cite{maruani2023voromesh} refines vertex positions using Voronoi diagrams, while PoNQ~\cite{maruani2024ponq} applies quadric error metrics for optimization. However, most of these approaches still depend on voxel-based representations, and they often fail to capture smooth surfaces and geometric features accurately due to the resolution constraint.
More recently, LMR~\cite{zhang2025high} proposes an adaptive meshing method that uses implicit surfaces and surface curvature to generate resolution-adaptive and lightweight meshes. However, it does not address anisotropic mesh representation from point clouds, so that there is still a gap between their adaptive meshing and the $L_2$ optimal approximation to a smooth surface, i.e., the anisotropy of triangles conforms to the eigenvalues and eigenvectors of the curvature tensors~\cite{Simpson:1994,Heckbert:1999}. 

\begin{figure*}[t]
\centering
\includegraphics[width=0.95\linewidth]{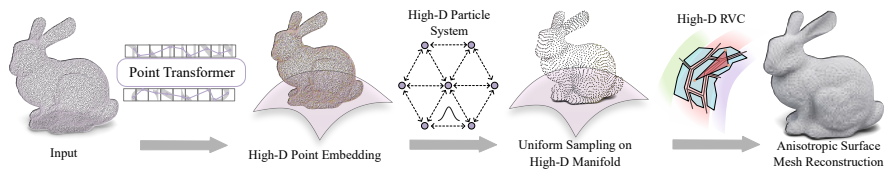}\vspace{-2mm}
\caption{The overview pipeline of the proposed neural high-d point embedding for anisotropic surface approximation (HD-PEA). Our method includes three main components: neural high-d Euclidean point embedding, high-d embedding manifold approximation by uniformly sparse sampling, and anisotropic manifold mesh reconstruction. 
}\vspace{-4mm}
\label{fig:pipeline}
\end{figure*}

This work addresses the above challenges in anisotropic surface mesh approximation from unstructured point clouds.
The \textit{main contributions} are:\vspace{-4mm}
\begin{itemize}
  \item Design a new computational framework to reconstruct high-quality anisotropic surface meshes from unstructured point clouds;
  \item Develop a scalable neural high-d Euclidean point embedding that maps large-scale point clouds onto a high-d manifold while preserving intrinsic Riemannian metrics, overcoming the mesh-dependent limitation of prior anisotropy losses, which prevents their direct application to point clouds;
  \item Develop a new tangent subspace approximation for the high-d embedding manifold to construct uniformly sparse samplings and reconstruct anisotropic meshes in the high-d space;
  \item Construct and release a variety of datasets for anisotropic surface mesh reconstruction and curvature tensor estimation along with the input point clouds.
\end{itemize}
The overview pipeline of the proposed neural high-d point embedding for anisotropic surface approximation (HD-PEA) is shown in Fig.~\ref{fig:pipeline}.

\section{Related Work}

\textbf{Neural Surface Meshing and Reconstruction.} In recent years, the deep learning-based approaches for meshing of shapes become popular in computer vision and geometric modeling. For instances, Point2Mesh~\cite{hanocka2020point2mesh}, Pixel2Mesh~\cite{wang2018pixel2mesh}, Pixel2Mesh++~\cite{wen2019pixel2mesh++} reconstruct meshes by deforming a template mesh to fit the input point cloud, but this typically restricts the output to the template's fixed topology. To address this, PointTriNet~\cite{sharp2020pointtrinet} directly predicts mesh connectivity by iteratively proposing and classifying triangle candidates. Although the method is local and differentiable, it often produces meshes with holes. NeuralMeshing~\cite{vetsch2022neuralmeshing} proposes differentiable meshing algorithm for extracting surface meshes from neural implicit representations, but it cannot guarantee watertightness and feature preservation. NMC~\cite{chen2021neural} and NDC~\cite{chen2022neural} improve the meshing templates of Marching Cubes (MC)~\cite{lorensen1998marching} and Dual Contouring (DC)~\cite{ju2002dual} by incorporating learned vertex predictions. VoroMesh~\cite{maruani2023voromesh} further refines vertex placement using Voronoi diagrams, and PoNQ~\cite{maruani2024ponq}  leverages a quadric error metric for optimization. Despite these advancements, most existing neural meshing techniques still rely on voxel-based representations, which constrain their accuracy and expressiveness, especially on low-resolution grids. To address this issue, most recently LMR~\cite{zhang2025high} proposes an adaptive Delaunay meshing that uses implicit surfaces and surface curvatures, but this method is not generalizable and efficient, as it requires optimizing each input point cloud individually. In summary, the existing methods do not address anisotropic mesh representation and how to extend them
into an arbitrary dimensional space is unexplored.

\hspace{-5.2mm}\textbf{High-Dimensional Geometric Embeddings.} Several approaches have been proposed to address specific classes of surface meshing and processing problems by embedding them in higher-dimensional spaces~\cite{canas2006surface,boissonnat2008anisotropic,kovacs2010anisotropic,levy2013variational,dassi2014curvature,dassi2015anisotropic}, compared with parameterization-based methods~\cite{chen2024neuraltps,noda2025learning} often struggling with complicated topologies due to their reliance on low-dimensional canonical domains. Their core idea is to transform an anisotropic meshing problem defined on a 3D surface into an equivalent isotropic meshing problem in 6D or a higher dimension. In particular, SIFHDE$^2$~\cite{Embedding2018} introduces an optimization-based approach to compute the high-d embedding space. NASM~\cite{Li2024} uses dot products between edge vectors on mesh faces to model anisotropy to learn the high-d embedding. However, since point clouds lack explicit tangent spaces and vector representations, such dot product-based losses cannot be directly applied. Moreover, the above methods are designed for mesh-based representations and cannot be directly applied to point clouds. Fundamentally, extending the Nash Embedding Theorem~\cite{Nash:1954,nash1956imbedding} to point clouds remains challenging, as existing formulations assume continuous manifolds and therefore cannot operate on discrete, unstructured point sets.

\hspace{-5.2mm}\textbf{Anisotropic Triangular Meshing.} Delaunay triangulation has been extended to anisotropic settings using two main approaches: refinement-based methods~\cite{borouchaki1997delaunayI,borouchaki1997delaunayII,dobrzynski2008anisotropic} and variational methods~\cite{chen2007optimal}. \cite{boissonnat2015anisotropic} employs a progressive vertex insertion strategy for anisotropic Delaunay refinement. \cite{rouxel2016discretized} approximates Riemannian Voronoi diagrams on surfaces using geodesic distance computations. \cite{fu2014anisotropic} proposes Locally Convex Triangulation (LCT), which aligns convex functions with Riemannian metrics for anisotropic meshing. \cite{budninskiy2016optimal} lifts points to convex functions to minimize reconstruction error, incorporating anisotropy, but their method is limited to anisotropies representable via convex functions. \cite{dai2024anisotropic} introduces a method for high-quality triangular meshing with prescribed anisotropy using metric-adapted embeddings based on cone singularities. Another group of anisotropic triangular meshing methods are based on computing the dual mesh from the anisotropic centroidal Voronoi tessellation (ACVT), such as~\cite{du2005anisotropic,valette2008generic,zhong2013}. All of these methods are designed for mesh-based representations and computations. Recently, \cite{zhong2019surface} develops a unified particle-based formulation for resamplings with specific patterns from original point clouds. However, it is not effective and scalable for a large-scale number of models, since it is a traditional model-based method and it also needs a curvature tensor for each model as input.\vspace{-4mm}

\section{Method}
\subsection{Neural High-D Euclidean Point Embedding}
\label{subsec:highd_embedding}
The Nash embedding theorem states that every smooth Riemannian manifold can be isometrically embedded into a high-d Euclidean space~\cite{Nash:1954,nash1956imbedding}. However, applying Nash Embedding Theorem to point clouds is challenging, as existing methods (e.g., SIFHDE$^2$~\cite{Embedding2018}, NASM~\cite{Li2024}) require continuous manifolds and cannot operate on unstructured points. We address this by introducing a barycentric-invariant isometric propagation that generates high-d embeddings from point clouds, by enforcing barycentric consistency, our method strictly satisfies the continuity and differentiability requirements of the embedding theorem, bridging Nash embedding to unstructured point clouds for the first time rather than continuous manifolds or mesh domains required by prior high-d embedding formulations. Motivated by this foundation, we propose a method to learn a high-d point embedding that preserves the intrinsic Riemannian geometry, given only its point cloud and normal vectors. 
This high-d point embedding is designed to approximate an isometric mapping of the underlying manifold which is the input for subsequent high-d embedding manifold approximation. 

\textbf{Data Generation for High-D Point Embedding.}
Existing mesh-based approaches~\cite{Embedding2018,dai2024anisotropic,Li2024} in geometry processing that consider surface anisotropy primarily operate on graph-like structures, where edges and connectivity are explicitly defined. These methods rely on well-structured meshes to model anisotropic behavior. In contrast, there is no existing method capable of generating ground truth high-d embeddings for unstructured point sets, particularly in a way that faithfully encodes manifold's local geometry (e.g., curvatures, tangent space, principal directions).

To address this gap, our data generation for point cloud embedding leverages the mapping from the original 3D Riemannian space to the target high-d Euclidean space. It can be described by an affine transformation~\cite{deformation_transfer,panozzo2014frame,Embedding2018}, where the transformation matrix inherently captures local manifold curvature and directional information. Since barycentric coordinates are preserved under affine transformations, we first sample points on the original 3D triangle surfaces and compute the barycentric coordinates of each sampled point with respect to its triangle. Using these same barycentric weights, we interpolate the corresponding high-d vertex embeddings to obtain the high-d coordinates of each sampled point. This process constructs a dense and high-quality ground truth point cloud in the target high-d manifold space, where it is both geometrically consistent and sensitive to local anisotropy. It provides a complete ground truth mapping from the 3D surface to its high-d embedding for supervised learning. Details are provided as follows.

\begin{wrapfigure}{r}{0.6\textwidth}\vspace{-8mm}
    \includegraphics[width=\linewidth]{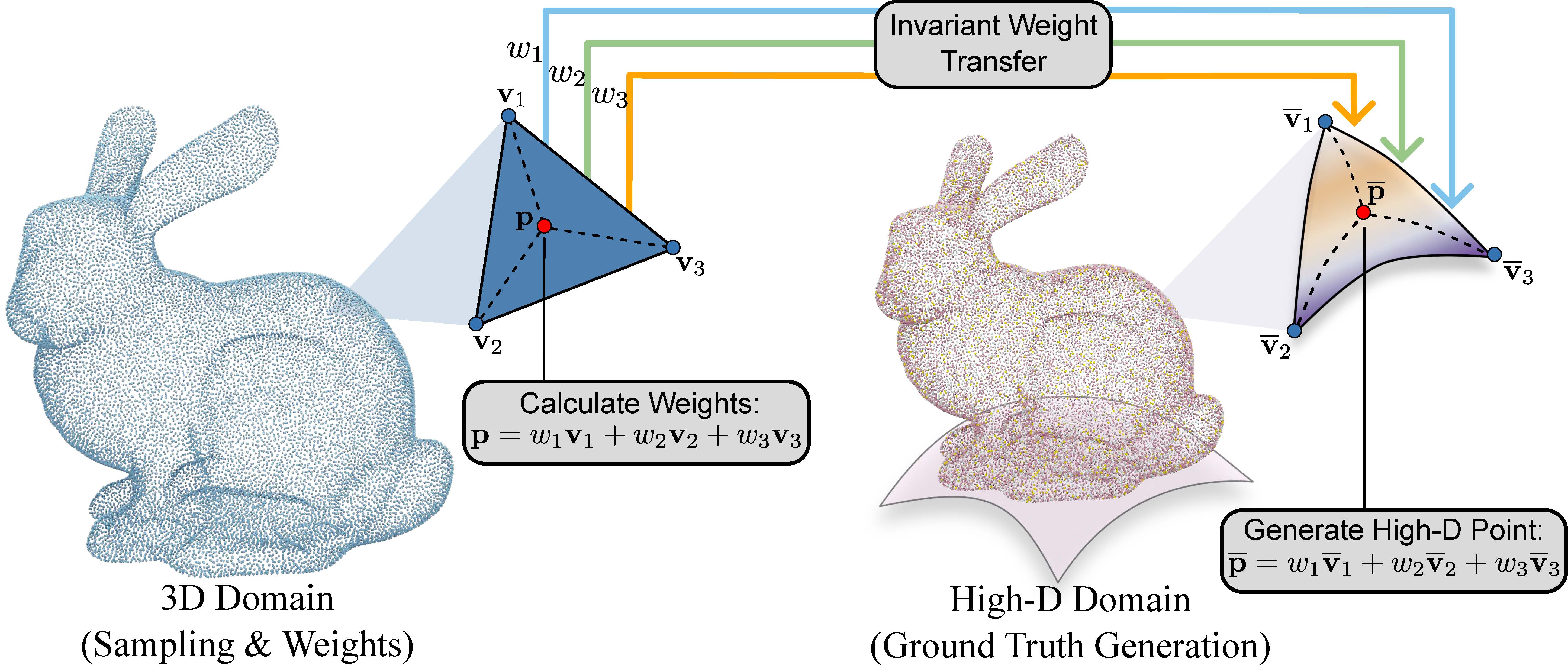}
    \caption{The procedure of data generation for high-d point embedding. The triangles represent the underlying mesh used in the invariant weight transfer.}\vspace{-8mm}
    \label{fig:barycentric}
\end{wrapfigure}

Formally, let the original 3D triangle mesh denotes as $\mathcal{M}=(\mathcal{V}, \mathcal{F})$, where $\mathcal{V}=\{\mathbf{v}_i\in \mathbb{R}^3 \}$ are the vertex positions and $\mathcal{F}$ is the set of triangle faces. The corresponding high-d embedded mesh is denoted as $\overline{\mathcal{M}} = (\overline{\mathcal{V}}, \overline{\mathcal{F}})$, where each vertex $\overline{\mathbf{v}}_i \in \mathbb{R}^{d}$ is the high-d embedding of $\mathbf{v}_i$ and $ d = 8$ (as suggested in~\cite{Embedding2018}), $\mathcal{F}=\overline{\mathcal{F}}$. As shown in Fig.~\ref{fig:barycentric}, for any point $\mathbf{p}$ sampled in a triangle face $\mathbf{f} = \{\mathbf{v}_1, \mathbf{v}_2, \mathbf{v}_3\}$, we compute its barycentric coordinates $(w_1, w_2, w_3)$ as:\vspace{-2mm}
\begin{align}
    \mathbf{p} = w_1\mathbf{v}_1+w_2\mathbf{v}_2+w_3\mathbf{v}_3,\\
    \text{where} \quad w_1+w_2+w_3=1, w_i\geq0.\vspace{-2mm}
\end{align}
The corresponding high-d embedding of $\mathbf{p}$, denoted $\overline{\mathbf{p}}\in \mathbb{R}^{d}$, is obtained by applying the invariant weight
transfer (i.e., same barycentric weights) to the high-d vertices:\vspace{-3mm}
\begin{align}
    \overline{\mathbf{p}} = w_1\overline{\mathbf{v}}_1+w_2\overline{\mathbf{v}}_2+w_3\overline{\mathbf{v}}_3.\vspace{-4mm}
\end{align}
By repeating this procedure across all triangles in $\mathcal{M}$, we obtain a dense set of point pairs ${(\mathbf{p}, \overline{\mathbf{p}})}$, where ${\mathbf{p}} \subset \mathbb{R}^3$ and ${\overline{\mathbf{p}}} \subset \mathbb{R}^{d}$. This provides a complete ground truth mapping from the 3D surface to its high-d Euclidean embedding, which can be used for our follow-up high-d point embedding learning task.

While the generated embeddings may partially inherit biases from the source meshes, our framework uses the mesh only during this data generation stage to maintain consistency with an underlying smooth Riemannian manifold. Unlike two-stage pipelines that suffer from intermediate reconstruction artifacts (discussed in Section~\ref{sec:evaluation}), our method learns the anisotropic metric field directly from raw point clouds in the high-d space.

\textbf{High-D Point Embedding Network and Loss.}
Formally, let $\mathbf{P}\in \mathbb{R}^{N\times3}$ denote a point cloud consisting of $N$ points in 3D space, and let $\mathbf{n}\in\mathbb{R}^{N\times3}$ represent the corresponding unit normal vectors. The input to our network is the concatenated representation $[\mathbf{P}, \mathbf{n}]\in \mathbb{R}^6$. A point-based transformer backbone $\mathcal{F}$, parameterized by $\theta_\mathcal{F}$, predicts the high-d coordinate extension $\tilde{\mathbf{P}}\in \mathbb{R}^{N\times(d-3)}$. The final high-d embedding is obtained by concatenating the original 3D coordinates with the predicted extension: $\overline{\mathbf{P}}= [\mathbf{P},\mathcal{F}([\mathbf{P},\mathbf{n}];\theta_\mathcal{F})]$. We apply Point Transformer V3~\cite{wu2024point} as our backbone architecture. To accommodate the required output dimensionality, we append a single linear layer to the end of the network.

To capture the local geometric structure of the underlying manifold from the input point clouds, we utilize the pairwise distances between neighboring points. Different from the prior work such as \cite{Li2024}, which models anisotropy using dot products between edge vectors defined on the given mesh faces, point clouds lack explicit tangent spaces and well-defined vector representations. These challenges make utilization of the dot product-based loss infeasible in our setting. Instead, we approximate the local geometry by pairwise distances, which implicitly encode the intrinsic Riemannian metric governing local lengths and angles. 

In prior works~\cite{lop_2007,wlop_2009,GPM_2018}, $L_1$-based losses tend to bias predictions toward the median of the target distribution, making them more effective for learning from diverse and non-uniform geometric structures. Building on this insight, we adopt Mean Absolute Error (MAE) between predicted and ground truth pairwise distances. This formulation promotes the preservation of relative geometric structure across diverse shapes and supports more accurate high-d embedding prediction. Our neighborhood distance-based loss is:\vspace{-2mm}
\begin{align}
    \mathcal{L}=\frac{1}{N}\sum_{i=1}^N\sum_{j\in\mathcal{N}(i)}|\left\| f(\mathbf{p}_i)-f(\mathbf{p}_j)\right\|-\left\| \overline{\mathbf{p}}_i-\overline{\mathbf{p}}_j  \right\||,\vspace{-6mm}
\end{align}
where $f(\mathbf{p}_i)$ denotes the predicted embedding coordinates of point $\mathbf{p}_i$, $\overline{\mathbf{p}}_i$ denotes the corresponding ground truth embedding coordinates. $\mathcal{N}(i)$ represents the set of neighboring points of $\mathbf{p}_i$. In our experiment, $\mathcal{N}(i)$ is set to be 40 and the embedding dimension is 8 as suggested in~\cite{Embedding2018}. This formulation encourages local geometric consistency between the prediction and the ground truth without relying on explicit edge connectivity. We demonstrate the effectiveness of our loss function in the ablation study. 

\textbf{Patch-Based Meta-Embedding Inference.}
Since the embedding network is trained on point clouds with a fixed number of points, applying it directly to larger point clouds can lead to degraded embeddings. To address this, we introduce patch-based meta-embedding inference (PMEI), i.e., an inference-time strategy that divides large point clouds into overlapping patches, processes each independently, and aligns the individual patch-based embeddings into a consistent global meta-embedding. It can efficiently handle arbitrarily large-scale point clouds without retraining while preserving embedding quality and consistency.


For a point cloud $\mathbf{P}\in\mathbb{R}^{N\times 3}$, we first construct a set of patch centers $\{\mathbf{c}_1,...,\mathbf{c}_l\}$ where $l<<N$ using farthest point sampling (FPS). For each $\mathbf{c}_i\in\{\mathbf{c}_1,...,\mathbf{c}_l\}$, we use its $k$ nearest neighbors to form a local patch $\mathbf{q}_i\in \mathbb{R}^{k\times 3}$, and the $k$ we choose guarantees the overlapping of the patches. Each patch $\mathbf{q}_i$ is then processed by the network to obtain its high-d Euclidean point embedding $\overline{\mathbf{q}}_i\in \mathbb{R}^d$. 
Specifically, for two overlapping patches $\overline{\mathbf{q}}_i$ and $\overline{\mathbf{q}}_j$ with shared domain $\mathcal{O}_{ij}$, the ideal consistency condition  $\overline{\mathbf{q}}_j|_{\mathcal{O}_{ij}}=\overline{\mathbf{q}}_i|_{\mathcal{O}_{ij}}$ is typically not satisfied. This inconsistency arises because standard attention mechanisms are not equivariant to affine transformations unless explicitly designed as~\cite{se3transformers2020}. Let $\{\mathbf{p}^k_i,\mathbf{p}^k_j\}_{k=1}^m$ denotes the set of $m$ corresponding point pairs between the overlapping regions $\overline{\mathbf{q}}_i|_{\mathcal{O}_{ij}}$ and $\overline{\mathbf{q}}_j|_{\mathcal{O}_{ij}}$. We formulate the patch-based embedding alignment as an optimization problem:\vspace{-3mm}
\begin{align}\label{eq:alignment}
T_{ij}^*=\arg\min\limits_{T_{ij}}\sum_{k=1}^m||\mathbf{p}^k_i-T(\mathbf{p}^k_j)||^2, \vspace{-6mm} 
\end{align}
where $T:\mathbb{R}^d\to \mathbb{R}^d$ is an affine transformation defined as $T(x)=\mathbf{A}\mathbf{x}+\mathbf{b}$ with $\mathbf{A}\in\mathbb{R}^{d\times d}$ and $\mathbf{b}\in \mathbb{R}^d$.

The embedding alignment process is formulated as a minimum spanning tree (MST) problem over the set of patch centers $\{\mathbf{c}_i\}^l_{i=1}$. Each center $\mathbf{c}_i$ is treated as a node, and the edge weight between two centers $\mathbf{c}_i$ and $\mathbf{c}_j$ is given by their Euclidean distance $||\mathbf{c}_i-\mathbf{c}_i||$. Starting from an arbitrary root center, the algorithm iteratively aligns the next nearest patch using Eq.~(\ref{eq:alignment}) and adds to the already aligned set, using a priority queue keyed by the inter-center distances-equivalent to Prim's algorithm for constructing an MST. This process ensures that the newly selected patches potentially have the largest overlapping region with the already aligned set, promoting more stable and consistent transformations. The PMEI algorithm and illustration are provided in Section H of Appendix.\vspace{-4mm}

\subsection{High-D Embedding Manifold Approximation}
\label{subsec:embed_app}
After obtaining the high-d point embedding, our goal is to obtain a piecewise linear approximation of the manifold $\mathcal{M}$ that follows a uniform sparse distribution in the high-d space. When this approximation is projected to original 3D manifold space, it naturally results in an anisotropic linear approximation. In this section, we present an approach for optimizing the vertex positions to achieve a uniform distribution in high-d poing embedding space. 

\textbf{Embedding Tangent Space Approximation.}
To operate on a point-based manifold in a high-d space, it is essential to have a guiding structure that constrains operations to lie on the manifold itself. Different from normals are used in 3D space~\cite{BoltchevaRVD2017,zhong2019surface} to define local orientation, we use the tangent subspace of the manifold to impose such constraints. The main reason is that if the surface is not a hypersurface (e.g., a 2D surface embedded in $\mathbb{R}^{8}$ in our work), the normal space becomes a subspace of dimension $d$, and there is no unique normal vector; instead, a normal subspace exists~\cite{sakai1996riemannian,zhang2004principal}. More generally, for an $m$-dimensional manifold $\mathcal{M}$ embedded in $\mathbb{R}^{d}$, the tangent space $\mathcal{T}_{\overline{\mathbf{p}}}\mathcal{M}$ at a point $\overline{\mathbf{p}}\in\mathcal{M}$ is a $m$-dimensional linear subspace of $\mathbb{R}^{d}$. It is spanned by linearly independent $m$ tangent vectors, which form the basis of $\mathcal{T}_{\overline{\mathbf{p}}}\mathcal{M}$. In our case, where the manifold is a 2D surface embedded in a high-d space, we estimate the local tangent space at each point $\overline{\mathbf{p}}\in \mathbb{R}^{d}$ using Principal Component Analysis (PCA)~\cite{pearson1901liii}. To accurately capture the intrinsic low-dimensional structure, PCA is applied to the $k$-nearest neighbors $\mathcal{N}_{\overline{\mathbf{p}}}=\{\overline{\mathbf{p}}_1,...,\overline{\mathbf{p}}_k\}$.

Let $\{\overline{\mathbf{p}}_i\}^k_{i=1}\subset \mathbb{R}^{d}$ denote the $k$-nearest neighbors of a center point $\overline{\mathbf{p}}\in\mathbb{R}^{d}$.
\begin{wrapfigure}{r}{0.25\textwidth}\vspace{-8mm}
\includegraphics[width=\linewidth]{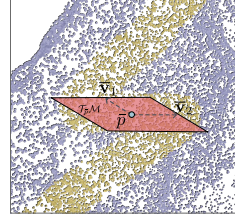}\vspace{-2mm}
\caption{The tangent space estimation in high-d space.}\vspace{-8mm}
\label{fig:tangent}
\end{wrapfigure}
The local centroid of these neighbors is given by $\hat{\overline{\mathbf{p}}}=\frac{1}{k}\sum^k_{i=1}\overline{\mathbf{p}}_i$. The local covariance matrix at $\overline{\mathbf{p}}$ is defined as $\mathbf{C}_{\overline{\mathbf{p}}}=\frac{1}{k}\sum_{i=1}^{k}(\overline{\mathbf{p}}_i-\hat{\overline{\mathbf{p}}})(\overline{\mathbf{p}}_i-\hat{\overline{\mathbf{p}}})^T$. The tangent basis at $\overline{\mathbf{p}}$ is then obtained by performing eigen decomposition of $\mathbf{C}_{\overline{\mathbf{p}}}$. For the target dimension $m$ of tangent subspace, the top $m$ eigenvectors $\overline{\mathbf{v}}$ of $\mathbf{C}_{\overline{\mathbf{p}}}$ corresponding to its largest $m$ eigenvalues, form an orthonormal basis: $\overline{\mathbf{V}} = [\overline{\mathbf{v}}_1,\overline{\mathbf{v}}_2,...,\overline{\mathbf{v}}_d]\in \mathbb{R}^{{d}\times m}$ for the estimated tangent space $\mathcal{T}_{\mathbf{\overline{p}}}\mathcal{M}\subset \mathbb{R}^{{d}\times m}$. In our setting, we use $d=8$ and $m=2$. Fig.~\ref{fig:tangent} illustrates the tangent space estimation in the high-d space. 


\textbf{Sampling Uniformity on Manifold.} Given a set of sparse seed points lying on the manifold of the target domain, our goal is to achieve an isotropic distribution over this manifold. Particle-based methods have been proven to be effective in low-dimensional settings (e.g., 2D or 3D)~\cite{zhong2013,zhong2019surface,Witkin1994} and have also been extended to high-d manifolds when an explicit mesh representation is available~\cite{Embedding2018}. However, their application to manifolds represented purely by point clouds in high-d spaces remains unexplored. To address this gap, we leverage the introduced tangent space approximation to constrain particle-based optimization directly to the local geometry of point-based high-d manifolds. More details about particle formulation and computations are discussed in Section A and Section B of Appendix. 
It is important to note that both  energy and forces are defined in terms of Euclidean distances between particles. However, the desired optimization domain is the target embedding manifold $\mathcal{M}$, rather than the entire ambient space $\mathbb{R}^{d}$. To constrain the optimization within the manifold, we use the tangent basis $\overline{\mathbf{V}}$ to project each particle back onto a local linear approximation of the embedding manifold. Specifically, for a given particle $\mathbf{\overline{x}}$, we first identify its nearest neighbors $\mathbf{\overline{x}}_p\in \mathcal{M}$ in the original 3D space, and perform the projection onto the tangent space $\mathcal{T}_{\mathbf{\overline{x}}_p}\mathcal{M}$ at $\mathbf{\overline{x}}_p$ as:
$\mathbf{\tilde{x}}_{\mathcal{T}_{\mathbf{\overline{x}}_p}\mathcal{M}} = \mathbf{\overline{x}}_p + \mathbf{\overline{V}} \mathbf{\overline{V}}^T(\mathbf{\overline{x}}-\mathbf{\overline{x}}_p)$, where $\overline{\mathbf{V}}\in \mathbb{R}^{d\times m}$ is the local orthonormal basis spanning the estimated tangent space $\mathcal{T}_{\mathbf{\overline{x}}_p}\mathcal{M}$. This projection ensures that particle's coordinate updates remain locally constrained to the manifold geometry in high-d space.

By minimizing the total energy subject to the constraint that all particles remain on the manifold $\mathcal{M}\subset \mathbb{R}^m$,  the system evolves toward a state of equilibrium where the forces acting on all particles are balanced. At equilibrium, this process yields a uniformly distributed and isotropic sampling over the target manifold.\vspace{-2mm}
\subsection{Anisotropic Manifold Reconstruction}

The next step is to compute a piecewise linear approximation of this manifold $\mathcal{M}$ by using the output seeds / particles $\mathbf{\tilde{x}}$ from the previous optimization, which is a set of well-distributed sparse seeds in $\mathbb{R}^d$ that lies on a manifold $\mathcal{M}$ of intrinsic dimension $m$, where $m<d$. . In our case, we assume that $\mathcal{M}$ is a smooth manifold with intrinsic dimension $m=2$. Inspired by ~\cite{BoltchevaRVD2017}, we adopt the concept of Tangential Delaunay Complex~\cite{boissonnat2010manifold,BOISSONNAT2004161,Freedman2002} to generalize the restricted Voronoi cells (RVCs) from 3D to a higher dimension, constructing our high-d tangential complex. Our method primarily operates within the $m$-dimensional tangent spaces at sampling points. Therefore, the computational complexity depends on the manifold dimension $m$ rather than high dimension $d$ in the ambient space.

\textbf{Manifold Tangent Space Approximation.} 
To reconstruct a piecewise linear approximation of the manifold in a higher dimension, we begin by forming a $m$-dimensional unit disk with a radius $r$ and uniformly discretizing into $n$ disks, where each radius is defined as $r=\alpha\cdot l_{\text{diag}}$, where $l_{\text{diag}}$ denotes the diagonal length of the shape's bounding box in high-d ambient space, and $\alpha$ is a user-defined scaling coefficient. Then, we estimate the tangent basis $\overline{\mathbf{V}}$ at each point $\mathbf{\overline{x}}_i\in \mathbb{R}^d$, using its nearest neighbors identified in the same space, and define an affine transformation $\mathbf{A}_{\mathbf{\overline{x}}}:\mathbb{R}^m\rightarrow\mathbb{R}^d$ that maps the intrinsic disk $\mathbf{D}^m \subset \mathbb{R}^m$ into the ambient space with dimension $d$. This mapping is: 
    $\mathbf{D}^d(\mathbf{\overline{x}}) = \overline{\mathbf{V}} \mathbf{D}^m + \mathbf{\overline{x}}$,
where $\mathbf{D}^d \in \mathbb{R}^d$ is an image of the disk in the ambient space centered at $\mathbf{\overline{x}}$. It forms a local linear approximation of the manifold around $\mathbf{\overline{x}}$.

\textbf{High-D RVC and Mesh Extraction.}
Once the tangent space is
\begin{wrapfigure}{l}{0.25\textwidth}\vspace{-8mm}
    \includegraphics[width=\linewidth]{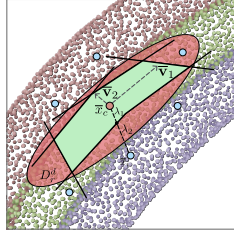}\vspace{-2mm}
    \caption{Illustration of a high-d RVC.}\vspace{-6mm}
    \label{fig:rvc}
\end{wrapfigure}
constructed at each seed, the restricted Voronoi cells (RVCs) forming the tangential complex are computed by clipping the embedded disk $\mathbf{D}^d$ with the bisectors between the seeds and their neighbors. An example is illustrated in Fig.~\ref{fig:rvc}. Since our target dimension is higher than 3, we apply re-entrant clipping method~\cite{reentrant1974}. Details are provided in Section C of Appendix.
As in~\cite{BoltchevaRVD2017,DEY2011483_cocone}, the result of the previous clipping step is a triangle soup. We adopt a manifold extraction method following~\cite{BoltchevaRVD2017}. Finally, we apply the post-processing step provided by Geogram~\cite{levy2015geogram} to fill any remaining holes from the previous step if necessary. Details are provided in Section D of Appendix. When the extracted mesh approximation is projected (via truncation) to original 3D manifold space, it naturally results in an anisotropic mesh approximation.\vspace{-4mm}

\section{Experiment}\vspace{-2mm}
We evaluate the effectiveness of our method in approximating surfaces by reconstructing high-fidelity lightweight anisotropic meshes and its applications from point clouds across a wide variety of shapes. Due to the page limit, the implementation details are provided in Section E of Appendix and some additional results are provided in Section G of Appendix. Detailed implementations of other comparison methods in the following section can be referred from their original papers. \textit{The source code and data will be publicly released after acceptance.}\vspace{-2mm}

\subsection{Datasets and Metrics}
\label{sec:dataset_metrics}
To thoroughly evaluate our method, we employ multiple datasets and applications that target different aspects of generalization and robustness. 

\textbf{Training Dataset.} We apply FPS to sample 40K points from each of the 240 selected models in the Thingi10K dataset~\cite{Thingi10K}. To mitigate biases arising from geometric variations in the training set, we apply data augmentation strategy followed by NASM~\cite{Li2024}. After applying this data augmentation strategy, the training set comprises 2,400 models of point clouds. The sampled point clouds are then normalized and estimated oriented normals using the Winding Number Normal Consistency (WNNC) method~\cite{Lin_WNNC}. The ground-truth high-d point embeddings are generated following Section~\ref{subsec:highd_embedding}. 


\textbf{Testing Datasets.} We conduct experiments on several challenging datasets, including Thingi10k dataset~\cite{Thingi10K}, Myles et al.'s dataset~\cite{Myles_2014} (i.e., 3D shapes from AIM@Shape and Stanford shape repository), TetWeave dataset~\cite{binninger2025tetweave}, and ScanNet dataset~\cite{dai2017scannet}. Thingi10k dataset serves as a baseline for evaluating our method. Others are unseen datasets, which are only used in the testing stage. All evaluations are conducted without any fine-tuning or retraining. As for point clouds acquisition, we utilize randomly sampling from surface meshes, synthetic scanning~\cite{blensor2011,huang2024surface}, noisy and irregular sampling, real scanning, etc. More details are provided in Section F of Appendix.

\textbf{Metrics.} Besides the qualitative evaluation and comparison, we provide quantitative evaluation. We adopt standard metrics for \textit{reconstructed surface accuracy}, including Chamfer distance (CD), F-Score (F1), normal consistency (NC), and Hausdorff distance (HD). To measure the \textit{anisotropic mesh quality} G, for each triangle $\triangle_{abc}$ in the final mesh, we use its approximated metric $\mathbf{Q}(\triangle_{abc}) = (\mathbf{Q}(\mathbf{x}_a) + \mathbf{Q}(\mathbf{x}_b) + \mathbf{Q}(\mathbf{x}_c))/3$, where $\mathbf{Q}(\cdot) = \sqrt{\mathbf{M}(\cdot)}$, to affine-transform it from the original anisotropic space into the Euclidean space. $\mathbf{M}(\cdot)$ is the curvature metric tensor. After that, we employ the isotropic triangular criteria~\cite{frey1999surface}, to evaluate the quality of generated anisotropic mesh as~\cite{zhong2013,Li2024}. The quality of a triangle is measured by $G_{\triangle_{abc}} = 2 \sqrt{3} \frac{S}{ph}$, where $S$ is the triangle area, $p$ is its half-perimeter, and $h$ is the length of its longest edge. G is the average qualities of all triangles. We report the \textit{average runtime} of all the methods.\vspace{-2mm}

\subsection{Evaluations}
\label{sec:evaluation}
\begin{figure}
    \centering
    \begin{minipage}[t]{0.48\textwidth}
        \includegraphics[width=\linewidth]{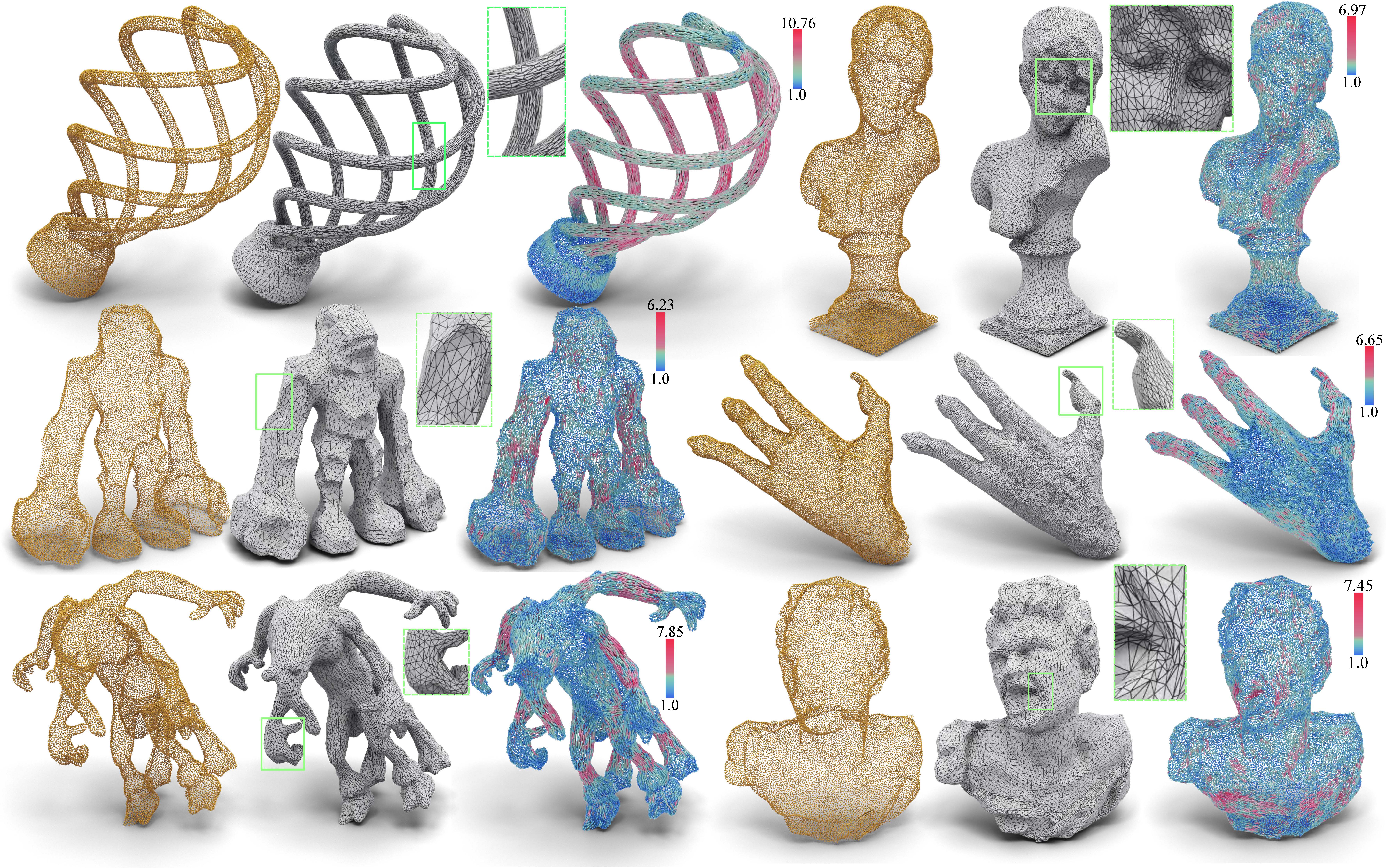}
        \caption{Our anisotropic surface mesh approximation from input point clouds. The models are selected from Thingi10K dataset~\cite{Thingi10K}. Left to right: input point clouds, reconstructed anisotropic meshes with zoom-in views, and estimated curvature tensors with stretching ratios for input point clouds in colors.}\vspace{-4mm}
        \label{ours}
    \end{minipage}
    \hspace{0.01\linewidth}
    \begin{minipage}[t]{0.48\textwidth}
        \centering
        \includegraphics[width=\linewidth]{fig/fig_compare_eccv-compressed.pdf}\vspace{-2mm}
        \caption{Our anisotropic surface mesh approximation results comparison with the state-of-the-art in surface mesh reconstruction. The models are selected from Myles et al.'s dataset~\cite{Myles_2014}. 
        NDC, POCO, and Poisson use grid size of $128^3$.}\vspace{-4mm}
        \label{mesh_compare}
    \end{minipage}
\end{figure}

\textbf{Baseline Sampling.}
We first evaluate our HD-PEA method on the testing set with 80 point cloud models selected from Thingi10K as shown in Fig.~\ref{ours}, and compare the results to several state-of-the-art in surface mesh reconstruction, including Poisson~\cite{kazhdan2006poisson}, NDC~\cite{chen2022neural}, POCO~\cite{boulch2022poco}, PoNQ~\cite{maruani2024ponq}, and LMR~\cite{zhang2025high} in Table~\ref{tab:comparison_Thingi10K}. We report both quantitative and qualitative results to demonstrate that HD-PEA achieves the best CD, F1, NC, and HD values. It justifies that incorporating anisotropic surface meshing strategy not only better captures geometric detail but also leads to more accurate surface approximations with fewer mesh elements. Compared with other deep learning-based methods, our approach requires only a few seconds for the entire computation.\vspace{-6mm}

\begin{table}[t]
    \centering
    \begin{minipage}[t]{0.48\textwidth}
        \caption{Quantitative comparison with our HD-PEA and state-of-the-art surface reconstructions on 80 models selected from Thingi10K dataset~\cite{Thingi10K}. The best results are highlighted in bold, and the second best results are underlined. Note: CD ($\times 10^{-5}$), HD ($\times 10^{-2}$), and Time (seconds). All metrics are averaged over 80 models.}\vspace{-10mm}
        \label{tab:comparison_Thingi10K}
        \begin{center}
        \resizebox{\linewidth}{!}{%
            \begin{tabular}{lllllllllll}
            \toprule
           Method & $\#V_{out}$ & $\#f_{out}$ & CD $\downarrow$ & F1 $\uparrow$ & NC $\uparrow$ & HD $\downarrow$ & Time \\
            \midrule
            Poisson & 38,346 & 76,684 &39.128& 0.912 & 0.970 & 1.821 & 19.271\\
            NDC($64^3$) & 4,134 & 8,327 & 0.798 & 0.986 & 0.978 & 0.921 & 4.971 \\
            NDC($128^3$) & 16,638 & 33,470 & \underline{0.676} & \underline{0.991} & \textbf{0.987} & \underline{0.716} & 5.532 \\
            POCO$(64^3$) & 8,237 & 16,497 & 29.817 & 0.853 & 0.951 & 6.521 & 4.367\\
            POCO($128^3$) & 33,590 & 67,196 & 33.107 & 0.888 & 0.960 & 6.862 & 15.588\\
             PoNQ   & 15,760 & 31,531 &11.729 & 0.967 & \underline{0.982} & 1.813 & 31.047\\
            LMR & 7,848 & 15,721 & 2.936 & 0.978 & 0.978 & 0.778 & 2,710.6 \\
            Ours &5,180 & 10,358 & \textbf{0.496} & \textbf{0.996} &\textbf{0.987} & \textbf{0.678} & 14.280 \\
            \bottomrule
            \end{tabular}%
        }\vspace{-6mm}
        \end{center}
    \end{minipage}
    \hspace{0.01\linewidth}
    \begin{minipage}[t]{0.48\textwidth}
        \caption{Quantitative comparison with our HD-PEA and state-of-the-art surface reconstructions on 75 synthetic scanned models selected from Myles et al.'s dataset~\cite{Myles_2014}. The best results are highlighted in bold, and the second best results are underlined. Note: CD ($\times 10^{-5}$), HD ($\times 10^{-2}$), and Time (seconds). All metrics are averaged over 75 models.}\vspace{-10mm}
        \label{tab:comparison_Myles_2014}
        \begin{center}
        \resizebox{\linewidth}{!}{%
            \begin{tabular}{llllllllll|}
            \toprule
           Method & $\#V_{out}$ & $\#f_{out}$ & CD $\downarrow$ & F1 $\uparrow$ & NC $\uparrow$ & HD $\downarrow$ & G $\uparrow$ & Time \\
            \midrule
            Poisson   & 77,296 & 154,594 & 26.346 & 0.922 & 0.966 & 2.147 & 0.528 & 39.298 \\
            NDC($64^3$) & 5,182 & 10,575 & 1.308 & 0.939 & 0.956 & 1.230 & 0.589 &5.276  \\
            NDC($128^3$) & 21,011 & 42,721 & \underline{0.749} & \underline{0.982} & 0.971 & \underline{0.928} & 0.588 & 5.928 \\
            \quad +NASM & 6,964 & 14,210 & 0.792 & 0.980 & \underline{0.975} & 1.265 & 0.576 & 24.701\\
            POCO$(64^3$) & 14,246 & 28,600 & 94.402 & 0.608 & 0.859 & 11.665 & 0.537 & 6.329 \\
            POCO($128^3$) & 58,871 & 117,837 & 90.462 & 0.626 & 0.875 & 11.601 & 0.530 & 22.869 \\
            PoNQ   & 24,685 & 49,368 & 7.610 & 0.963 & 0.973 & 1.511 & 0.596 & 35.834 \\
            NKSR   & 12,023 & 24,040 & 1.461 & 0.954 & 0.966 & 1.745 & 0.525 & 0.705\\
            SAP & 31,692 & 63,320 & 2.081 & 0.976 & 0.972 & 1.447 & 0.526 & 6.575  \\
            SIREN & 696,845 & 1,393,631 & 58.96 & 0.843 & 0.953 & 3.593 & 0.522 & 4658.97\\
            LMR & 7,590 & 15,124 & 10.048 & 0.964 & 0.972 & 2.471 & \underline{0.629} & 3,153.8\\
            Ours & 7,028 & 13,989 & \textbf{0.649} & \textbf{0.989} & \textbf{0.979} & \textbf{0.798} & \textbf{0.702} & 14.594\\
            \bottomrule
            \end{tabular}%
        }\vspace{-2mm}
        \end{center}
    
    \end{minipage}
\end{table}

For Marching Cubes-like methods, such as Poisson, NDC, and POCO, often exhibit undersampling artifacts such as zigzag patterns, due to their inflexible sampling directions in Fig.~\ref{mesh_compare}. Although PoNQ employs QEM~\cite{garland1997surface} for surface approximation, it still relies on Delaunay triangulation, which typically produces more vertices than desired. LMR (one of the most recent adaptive mesh reconstruction methods) conforms to surface curvatures by using adaptive meshes; however, it does not adequately control stretching ratios for anisotropic triangles, leading to noticeable number of degenerate triangles. Another limitation lies in the reliance of LMR on an optimization-based process that is computationally expensive, thereby limiting its practicality. The detailed runtime is shown in Table~\ref{tab:comparison_Thingi10K}.

\textbf{Synthetic Scanning.}
To further justify capability of our method, this experiment evaluates our method on synthetic scanned point clouds without any fine-tuning or retraining. 
We evaluate our HD-PEA method on 75 synthetic scanned point clouds generated from Myles et al.'s dataset using Blensor~\cite{blensor2011} as in~\cite{huang2024surface}. For each model, we simulate realistic scanning on point clouds containing 40K points. To provide a more comprehensive evaluation under these realistic scanning conditions, we expand our baselines from the previous subsection to include three additional state-of-the-art methods: NKSR~\cite{huang2023nksr}, SAP~\cite{Peng2021SAP}, and SIREN~\cite{sitzmann2019siren}. Fig.~\ref{mesh_compare} and Table~\ref{tab:comparison_Myles_2014} show that our method consistently achieves the best qualitative and quantitative performance with lightweight mesh elements.

\begin{figure}[t]
    \centering
    \begin{minipage}[t]{0.28\textwidth}
        \includegraphics[width=\linewidth]{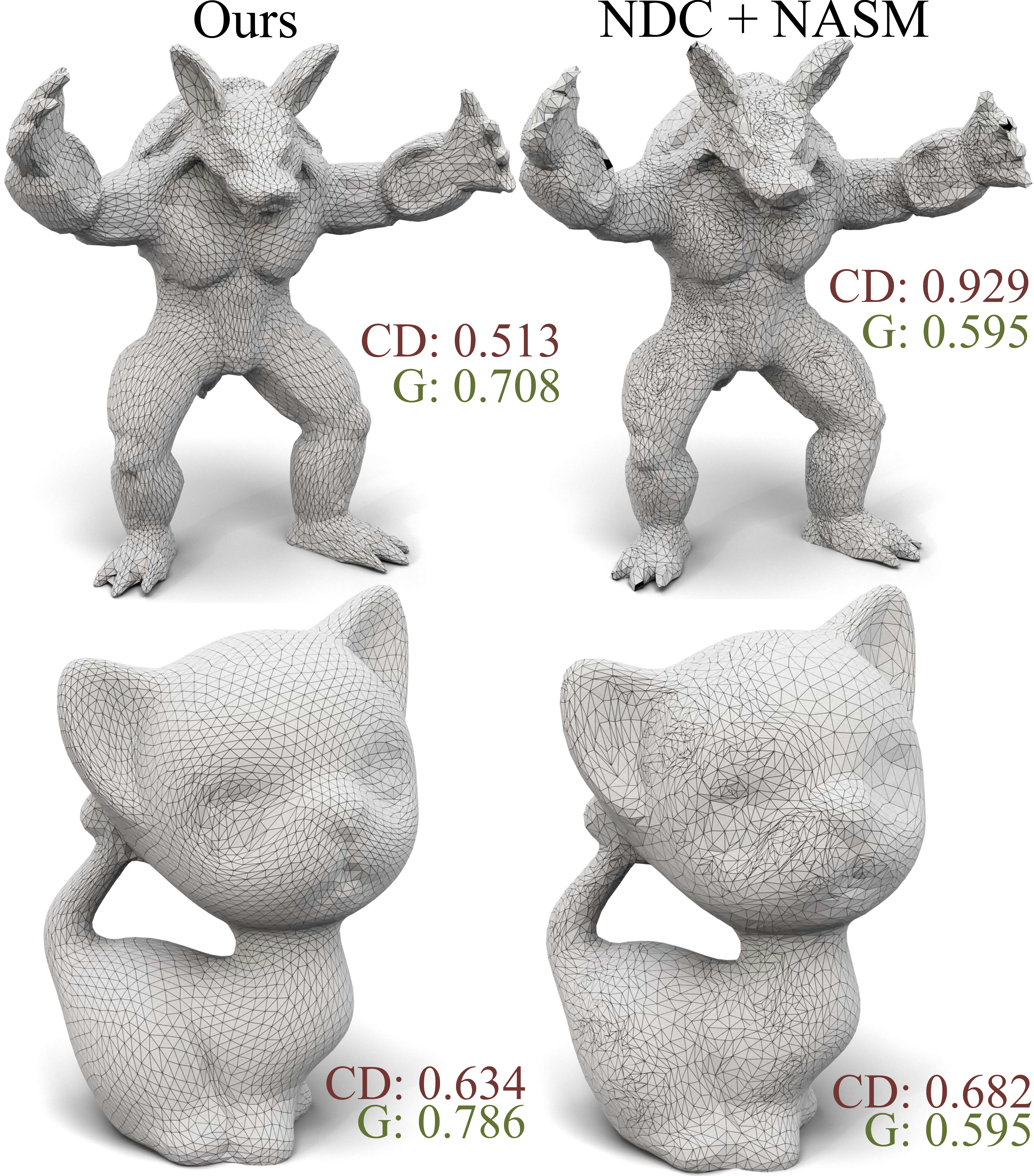}
        \caption{Comparison between ours and NDC+NASM, where HD-PEA is better on CD$\downarrow$ and G$\uparrow$.}\vspace{-4mm}
        \label{fig:ndc_nasm}
    \end{minipage}
    \hspace{0.01\linewidth}
    \begin{minipage}[t]{0.68\textwidth}
        \centering
        \includegraphics[width=\linewidth]{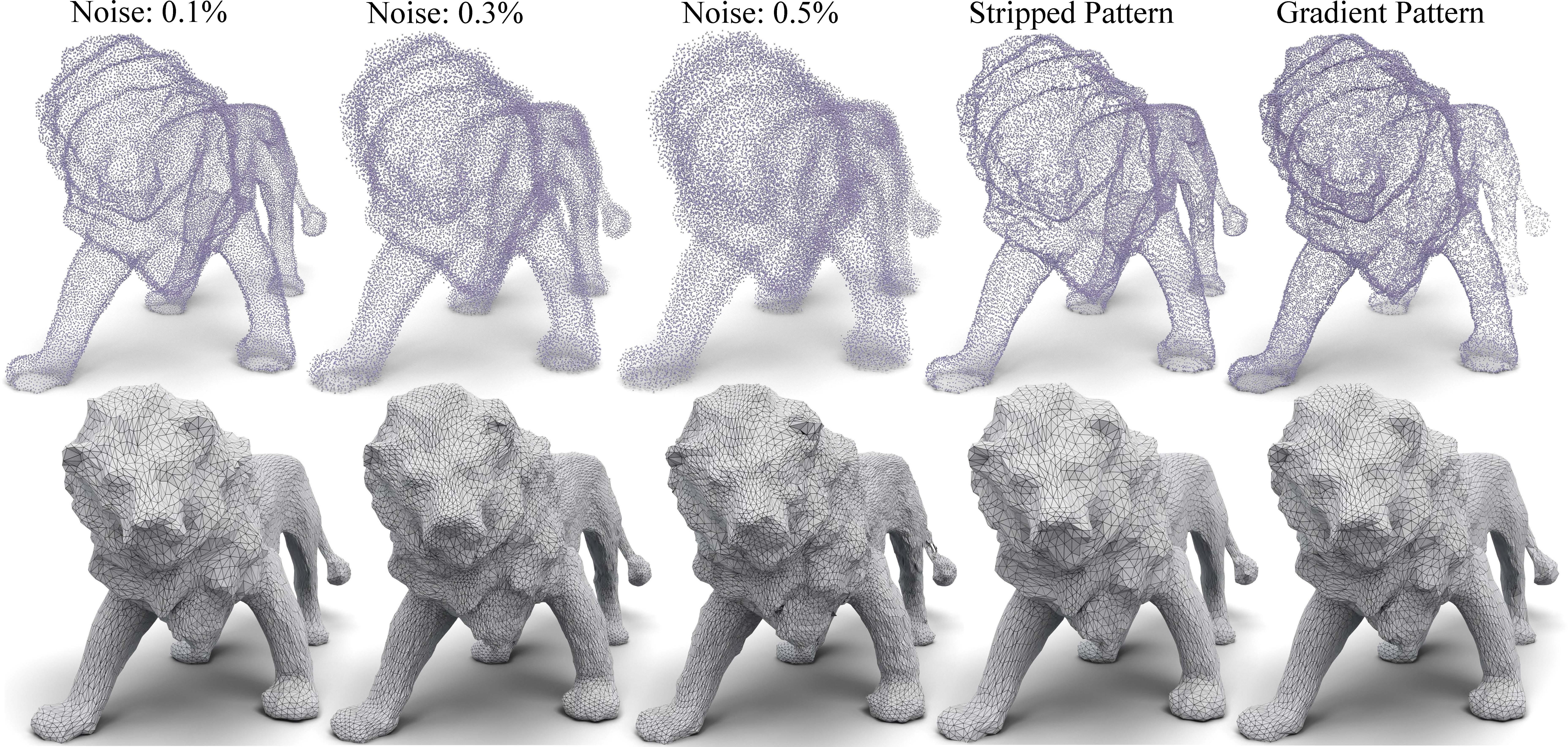}\vspace{-2mm}
        \caption{Our anisotropic surface mesh approximation results on a Lion model from input point clouds with different noise levels and irregular sampling patterns.}\vspace{-4mm}
        \label{noise_compare}
    \end{minipage}
\end{figure}

Although our method and NASM (one of the most recent anisotropic meshing methods) are inspired by the Nash Embedding Theorem, target fundamentally different problems. HD-PEA reconstructs anisotropic surfaces directly from raw point clouds, while NASM performs remeshing on an existing mesh. HD-PEA operates on raw point clouds without any intermediate mesh. In practice (see Fig.~\ref{fig:ndc_nasm}), we use NDC($128^3$) to generate the intermediate mesh for NASM, as NDC performed is the second-best method (besides ours) in our previous experiments. However, such two-stage pipelines (e.g., NDC+NASM) are sensitive to the quality of the intermediate mesh. As shown in Fig.~\ref{mesh_compare}, NDC produces severe aliasing artifacts, which subsequently cause NASM to overfit. By learning the anisotropic metric field directly in high-d embedding space from raw points, our HD-PEA avoids this artifact-prone stage, resulting in more stable and faithful reconstructions with consistently better surface accuracy (CD, F1, NC, HD) and anisotropic mesh quality metric G.

\textbf{Challenging Scanning.} 
In this section, we evaluate the performance of our method on degraded point clouds to demonstrate its robustness and practical applicability. Raw point clouds acquired from 3D scanners or multi-view reconstructions often suffer from measurement noise, outliers, and uneven sampling densities. To simulate these challenging conditions, we build upon the synthetic scanned dataset introduced in the previous section and apply some perturbations. Specifically, we add Gaussian noise with standard deviations of 0.001, 0.003, and 0.005 to simulate three levels of point-wise perturbations. Following the standard practice, we incorporate a lightweight preprocessing using the Moving Least Squares (MLS) method~\cite{mls_tog_2005}. MLS performs local surface fitting to smooth point clouds, effectively reducing high-frequency noise while preserving fine geometric details. This denoising step is applied prior to inference and adds minimal computational overhead.


To simulate irregular sampling density, we resample scanned point clouds using striped and gradient density variations, such that certain regions of the surface are more sparsely sampled, followed by PCP-Net~\cite{GuerreroEtAl:PCPNet:EG:2018}. Fig.~\ref{noise_compare} and Table~\ref{tab:mesh_input} demonstrate the good robustness of our method.

\textbf{Complicated Geometry and Large-Scale Scene.} We further evaluate our method on point clouds with complex geometry details and scene-level datasets containing a significantly large number of points. In this setting, we employ the proposed patch-based meta-embedding inference (PMEI) to demonstrate the scalability and adaptability to large-scale data. This evaluation covers both model-level shapes with fine geometric details (from Myles et al.'s dataset) and scene-level point clouds (ScanNet dataset~\cite{dai2017scannet}) with hundreds of thousands of input points. Figure~\ref{fig:scannet} demonstrates that our method has strong performance on large-scale point clouds on scene levels, even without fine-tuning or retraining. We also include both qualitative and quantitative comparisons between our method and NDC. Our method achieves a lower CD than NDC (0.786 vs. 0.814), while using only less than one-third of the mesh resolution. More results can be found in Section G of Appendix.



\begin{figure}[t]
\centering
\includegraphics[width=0.95\linewidth]{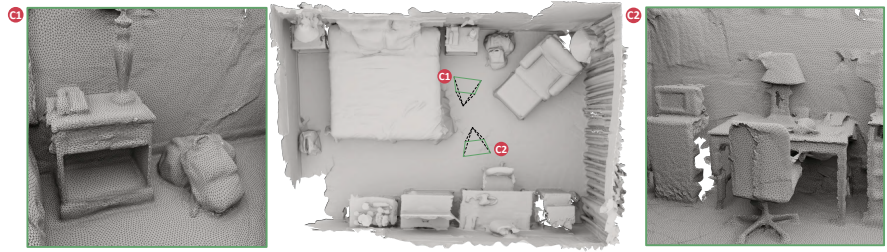}\vspace{0mm}
\caption{Our reconstruction of a ScanNet scene (some holes remain due to missing regions in the scanned input). Portions of the walls and roof are removed to reveal the interior geometry.
}\vspace{-6mm}
\label{fig:scannet}
\end{figure}

\begin{table}
    \centering
    \begin{minipage}[t]{0.48\textwidth}\vspace{-4mm}
        \caption{Quantitative evaluation on our HD-PEA with different levels of noise on samplings as well as irregular sampling schemes on 75 synthetic scanned point clouds from Myles et al.'s dataset. Note: CD ($\times 10^{-5}$), HD ($\times 10^{-2}$), and Time (seconds). All metrics are averaged over 75 models.}\vspace{-8mm}
        \label{tab:mesh_input}
        \begin{center}
        \resizebox{\textwidth}{!}{%
            \begin{tabular}{llllllllll}
            \toprule
           Method & $\#V_{out}$ & $\#f_{out}$ & CD $\downarrow$ & F1 $\uparrow$ & NC $\uparrow$ & HD $\downarrow$ & Time \\
            \midrule
            real scan+0.1\%  & 8,684 & 17,255 & 0.822 & 0.978 & 0.974 & 0.893 & 14.809 \\
            real scan+0.3\%  & 9,341 & 18,521 & 1.000 & 0.966 & 0.959 & 1.125 & 14.973 \\
            real scan+0.5\%  & 9,601 & 19,043 & 1.417 & 0.933 & 0.947 & 1.358 & 14.997\\
            gradient  & 7,526 & 14,907 & 0.730 & 0.982 & 0.976 & 0.970 & 14.630\\
            striped  & 7,186 & 14,296 & 0.688 & 0.985 & 0.978 & 0.894 & 14.597\\
            \bottomrule
            \end{tabular}%
        }\vspace{-6mm}
        \end{center}
    \end{minipage}
    \hspace{0.01\linewidth}
    \begin{minipage}[t]{0.48\textwidth}\vspace{-4mm}
        \caption{Ablation study on different loss functions for neural high-d point embedding on 80 models selected from Thingi10K dataset. The best results are highlighted in bold. Note: CD ($\times 10^{-5}$) and HD ($\times 10^{-2}$). All metrics are averaged over 80 models.}\vspace{-8mm} 
        \label{tab:ablation}
        \begin{center}
        \resizebox{\textwidth}{!}{%
            \begin{tabular}{lllllllllllll}
            \toprule
           Method & $\#V_{out}$ & $\#f_{out}$ & CD $\downarrow$ & F1 $\uparrow$ & NC $\uparrow$ & HD $\downarrow$  \\
            \midrule
            MSE   & 12,632 & 48,368 & 30.864 & 0.953 & 0.949 & 2.148 \\
            MAE   & 12,116 & 48,421 & 29.737 & 0.958 & 0.952 & 2.044 \\
            NDHDE w/ MSE & 8,688 & 48,991 & 30.072 & 0.984 & 0.979 & 1.065 \\
            NDHDE w/ MAE (Ours) & \textbf{5,180} & \textbf{10,358} & \textbf{0.496} & \textbf{0.996} &\textbf{0.987} & \textbf{0.678}\\
            \bottomrule
            \end{tabular}%
        }\vspace{-6mm}
        \end{center}
    \end{minipage}
\end{table}



\textbf{Curvature Tensor Estimation.}
One downstream application of our neural high-d point embedding is to estimate the curvature tensors from the input point clouds. For each point $\mathbf{p}$ in the original 3D space, we identify its the six nearest neighbors $\mathbf{p}_j$ and construct a matrix $\mathbf{A}\in \mathbb{R}^{6\times3}$, where each row corresponds the vector $\mathbf{p}-\mathbf{p}_j$. We also get the $\overline{\mathbf{p}}_j$ which are the corresponding predicted high-d embeddings of these 3D neighbors and construct a matrix $\mathbf{B}\in \mathbb{R}^{6\times 8}$ where each row represents the vector $\overline{\mathbf{p}}-\overline{\mathbf{p}}_j$. We then solve the linear system of $\mathbf{A}\mathbf{x}=\mathbf{B}$ in least squares sense, yielding a matrix $\mathbf{x}\in\mathbb{R}^{3\times 8}$. Applying singular value decomposition (SVD) to $\mathbf{x}$, we extract the three principal directions from its left singular vectors as the direction of the principal curvatures. The anisotropy stretching ratio is defined as $s_1/s_2$ where $s_1$ and $s_2$ denote the second and third singular values, respectively. Fig.~\ref{ours} illustrates the estimated curvature tensors. We visualize each tensor using ellipsoid-style glyphs, where the axes represent the principal curvature directions. The color encodes the anisotropy, defined by the stretching ratios of the tensors. Another application of our HD-PEA method is to analyze its ability to preserve surface curvature characteristics, which directly influence the rendering quality of the surface roughness or smoothness as shown in Section G of Appendix.\vspace{-2mm}


\subsection{Ablation Study}
\begin{figure}
    \centering
    \begin{minipage}[t]{0.48\textwidth}
        \includegraphics[width=\textwidth]{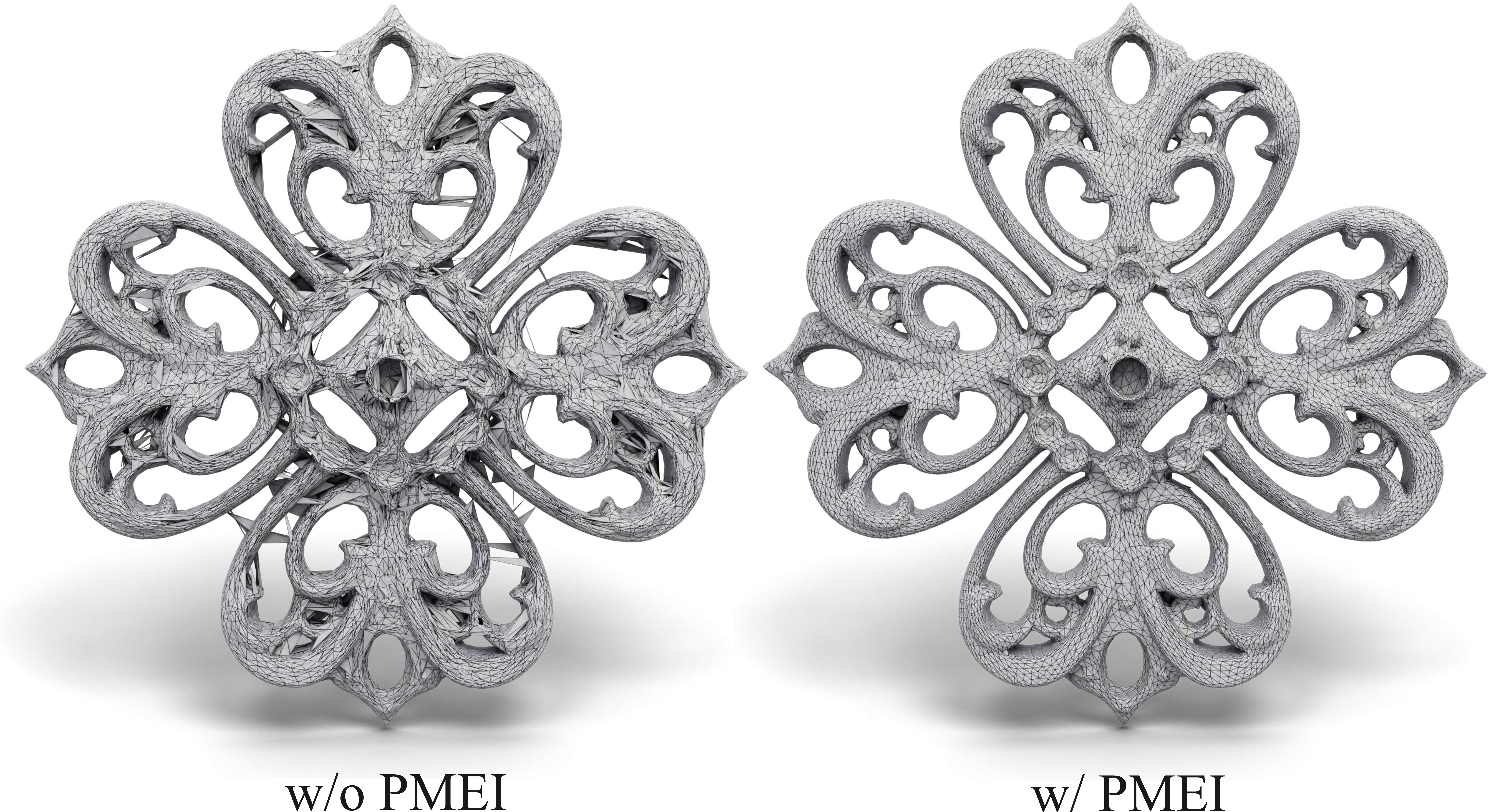}\vspace{-2mm}
        \caption{Comparison w/o and w/ patch-based meta-embedding inference (PMEI) on a complicated geometric model.}\vspace{-4mm}
        \label{fig:PSEI_compare}
    \end{minipage}
    \hspace{0.01\linewidth}
    \begin{minipage}[t]{0.48\textwidth}
        \includegraphics[width=\textwidth]{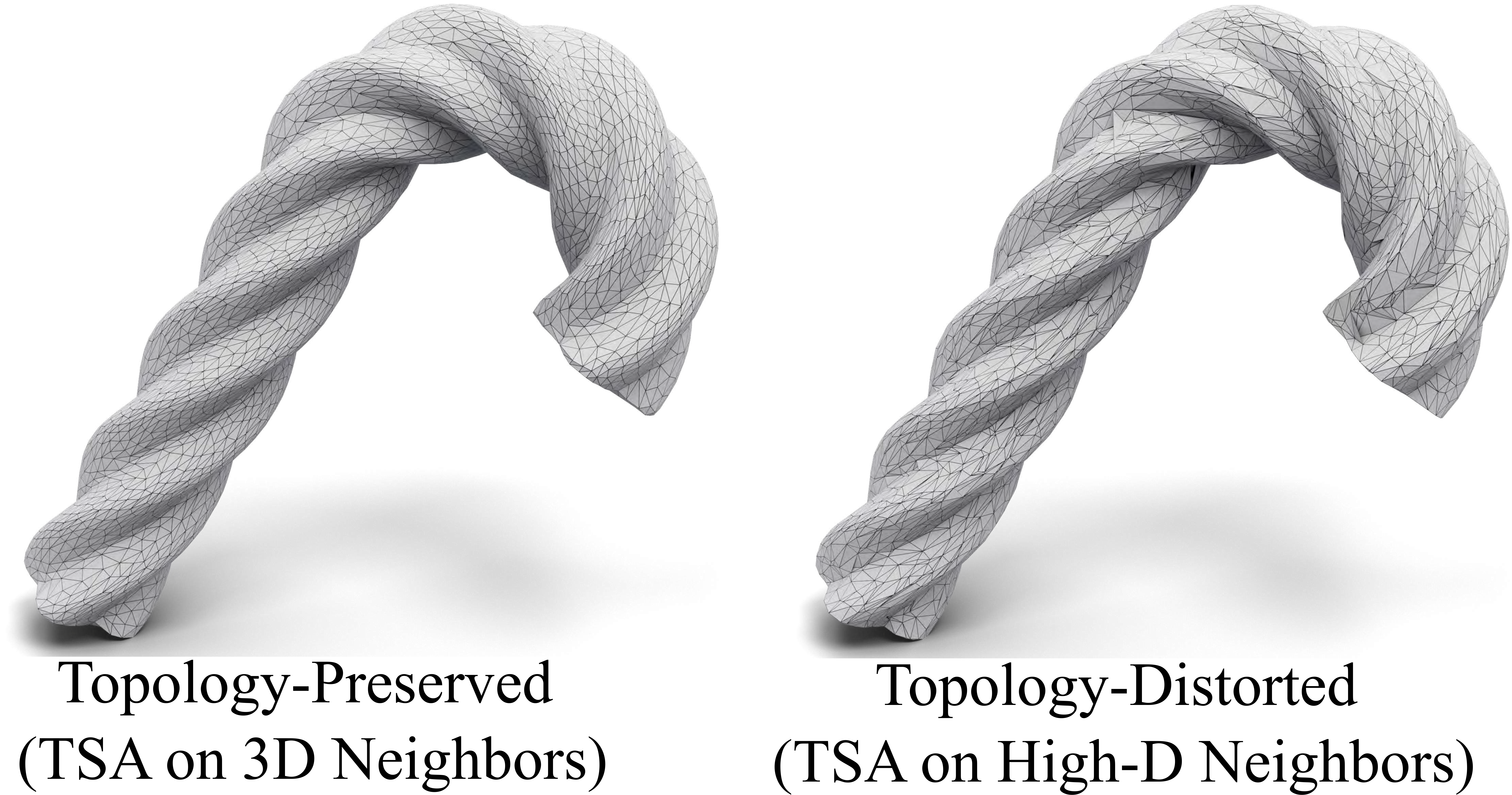}\vspace{-2mm}
        \caption{Comparison on the proposed embedding tangent space approximation (TSA) with 3D and high-d neighbors.}\vspace{-4mm}
        \label{fig:ablation_pca}
    \end{minipage}
\end{figure}

To demonstrate the effectiveness of our proposed neighborhood distance-based high-d embedding loss, i.e., NDHDE w/ MAE (in Section~\ref{subsec:highd_embedding}), we perform an ablation study comparing it against two commonly-used losses, i.e., Mean Squared Error (MSE) and Mean Absolute Error (MAE), both of which compute point-wise differences between predicted and ground truth embeddings. We also evaluate an MSE-based NDHDE and the ablation study is performed on Thingi10k dataset. Table~\ref{tab:ablation} shows that our proposed NDHDE w/ MAE loss is most effective on the anisotropic surface mesh reconstruction w.r.t. all surface accuracy metrics. Furthermore, we also show the necessity of applying the proposed PMEI (in Section~\ref{subsec:highd_embedding}) on large-scale complicated data in Fig.~\ref{fig:PSEI_compare}. Fig.~\ref{fig:ablation_pca} shows that applying PCA to 3D neighbors for embedding tangent space approximation and particle optimization constraints (in Section~\ref{subsec:embed_app}) preserves topology better than using high-d neighbors.\vspace{0mm}

\subsection{Failure Case}

\begin{figure}[h]\vspace{-2mm}
\centering
\includegraphics[width=0.6\textwidth]{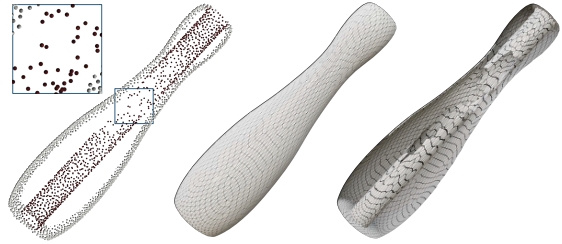}\vspace{-2mm}
\caption{Failure case. Left: input point cloud; Middle: our result from the surface view; Right: our result from the interior view. 
}\vspace{-8mm}
\label{failure_case}
\end{figure}

Figure~\ref{failure_case} illustrates a failure case on a challenging geometric shape with incomplete input point clouds from practical scanning. The clipped view removes front- and back-facing points to better highlight the severely missing interior samples. Such incomplete sampling makes neural high-d point embedding and manifold approximation particularly challenging for thin structures (e.g., the slender cylinder), resulting in invalid or missing mesh elements in these regions.\vspace{-4mm}

\section{Conclusion}\vspace{-2mm}
In this paper, we introduce a new scalable anisotropic surface approximation method through the high-d Euclidean point embedding. To our knowledge, this is the first deep learning-based method capable of generating high-quality and high-fidelity anisotropic surface meshes from unstructured point clouds without requiring a pre-computed mesh or metric field. Compared with existing surface mesh reconstruction methods, our HD-PEA achieves significantly better performance in surface accuracy, mesh quality, lightweightness, and scalability. Moreover, it shows strong potential for applications in robust large-scale surface reconstruction, surface rendering, and curvature tensor estimation.\vspace{6mm}

\hspace{-5.2mm}\textbf{Acknowledgments.} The authors would like to thank the anonymous reviewers for their valuable comments and suggestions. Hongbo Li, Haikuan Zhu, Jing Hua, and Zichun Zhong were partially supported by National Science Foundation (OAC-1910469 and OAC-2311245), National Institutes of Health (R61NS119434), and General Motors research grant.


%
%
\bibliographystyle{splncs04}
\bibliography{references}

\clearpage
\appendix

\section*{\Large Appendix: Learning Manifolds in High-D Point
Embedding for Anisotropic Surface
Approx}

\section{High-D Particle Formulation}
Given $n$ seed points $\mathbf{\overline{X}} = \{\mathbf{\overline{x}}_i\in\mathbb{R}^{d}|i=1,...,n\}$, lying on the manifold $\mathcal{M}\subset \mathbb{R}^{d}$, where ${d}$ is the dimension of the high-d ambient space. In our case, this manifold is represented by a set of points $\overline{\mathbf{P}} =\{\overline{\mathbf{p}}_i\in \mathbb{R}^{d}|i=1,...,N\}$. The inter-particle energy between particles $i$ and $j$ is defined as:\vspace{-1mm}
\begin{align}
\label{particle_energy}
    E^{ij}_p = e^{-\frac{||\mathbf{\overline{X}}_i-\mathbf{\overline{X}}_j||^2}{4\sigma^2}}.
\end{align}
Here $\sigma$, called kernel width, is the fixed standard deviation of the Gaussian kernels. 
More details about $\sigma$ are discussed in Section B of Appendix. Clearly, $E^{ij}_p=E^{ji}_p$.

The gradient of $E^{ij}$ w.r.t. $x_j$ can be interpreted as the force $\mathbf{F}^{ij}$ exerted on particle $j$ by particle $i$:
\begin{align}
\label{particle_force}
    \mathbf{F}^{ij}=\frac{\partial E^{ij}}{\partial \mathbf{\overline{x}}_j}=\frac{(\mathbf{\overline{x}}_i-\mathbf{\overline{x}}_j)}{2\sigma^2}e^{-\frac{||\mathbf{\overline{x}}_i-\mathbf{\overline{x}}_j||^2}{4\sigma^2}}.
\end{align}
According to Newton's third law of motion, the force satisfies $\mathbf{F}^{ij}=-\mathbf{F}_{ji}$. 

In our implementation, we leverage automatic differentiation to compute the forces by backpropagating through the explicit energy formulation in Eq.~(\ref{particle_energy}). Therefore, the optimization function is formulated as:
\begin{align}
\min_{\{\mathbf{\overline{x}}_i\}} \quad & E = \sum_i \sum_{j \neq i} E^{ij}, \hspace{4mm} \text{subject to} \hspace{2mm} \mathbf{\overline{x}}_i \in \mathcal{M},\  \forall i.
\end{align}

\section{Discussion about Kernel Width}
The Gaussian kernel energy defined in Eq.~(\ref{particle_energy}), and the corresponding force in Eq.~(\ref{particle_force}) with the kernel width $\sigma$, determine the effective area of influence for each seed point. The gradient of the energy reaches its maximum when the distance between particles is approximately $\sigma$ and approaches zero when the distance is much smaller or much larger than $\sigma$. If $\sigma$ is set too small, the Gaussian kernels barely overlap, resulting in negligible interactions between particles. Conversely, if $\sigma$ is set too large,  nearby particles are pushed toward the same location, causing excessive overlap where multiple particles converge onto a single point. To balance these effects, we set $\sigma$ proportional to the average kernel “radius” under the assumption of a uniform distribution over the target surface: $\overline{\Omega}: \sigma =c \sqrt{|\overline{\Omega}|/n}$, where $c$ is a user-defined scaling constant, $|\overline{\Omega}|$ is  the total surface area of the manifold, and $n$ is the number of seed points.

To estimate the surface area $\overline{\Omega}$ from a point cloud, we approximate the local surface area around each point $p_i$ using its nearest neighbors, i.e., typically the 12 nearest neighbors in the original 3D space. The radius of the local disk is computed as the average distance $\overline{d}_i$ between $p_i$ and its 12 neighbors. The local surface area associated with $p_i$ is then approximated as $\overline{S}_i=\alpha\pi\overline{d}_i^2$, where $\alpha$ is a scaling factor that compensates for local overlap and coverage. Summing the local areas over all points yields an estimate of the total surface area $\overline{\Omega}$ represented by the point clouds, i.e., $|\overline{\Omega}| = \sum^N_i \overline{S}_i$. In our experiments, we set $c=0.32$ and $\alpha=1.3$.

\section{Computing RVC Through High-D Clipping}
Once the tangent space is constructed at each seed, the restricted Voronoi cells (RVCs) forming the tangential complex are computed by clipping the embedded disk $\mathbf{D}^d$ with the bisectors between the seeds and their neighbors. An example is illustrated in Fig. 4 of the main paper. Since our target dimension is higher than 3, we apply re-entrant clipping~\cite{reentrant1974} to handle the high-d clipping problem as follows. 

Given the bisector $\mathbf{B}[\mathbf{\overline{x}}_0, \mathbf{\overline{x}}_1]$ and a cell's edge with two endpoints $\mathbf{a}$ and $\mathbf{b}$ that intersect the bisector, the intersection $\mathbf{I}$ can be obtained by:\vspace{-2mm}
\begin{align}
    \mathbf{I} = \lambda_1 \mathbf{a}+\lambda_2 \mathbf{b},\hspace{4mm}
    \text{where} 
    \begin{cases}
        \lambda_1 = \frac{l_1}{|\mathbf{a}\mathbf{b}|} \\
        \lambda_2 = \frac{l_2}{|\mathbf{a}\mathbf{b}|} 
    \end{cases},
\end{align}
where $l_1$ is the perpendicular distance from $\mathbf{a}$ to bisector $\mathbf{B}[\mathbf{\overline{x}}_0,\mathbf{\overline{x}}_1]$, $l_2$ is the perpendicular distance from $\mathbf{b}$ to bisector $\mathbf{B}[\mathbf{\overline{x}}_0,\mathbf{\overline{x}}_1]$. To calculate the perpendicular distance, we can calculate the distance between two parallel planes. The bisector can be defined as $d=-\frac{1}{2}(\mathbf{\overline{x}}_0+\mathbf{\overline{x}}_1)\cdot \overrightarrow{\mathbf{n}}$, and $\overrightarrow{\mathbf{n}}$ is the direction of $\overrightarrow{\mathbf{\overline{x}}_0\mathbf{\overline{x}}_1}$. The plane which passes through endpoint $\mathbf{a}$ and is parallel to the bisector is $d_1=-{\mathbf{a}}\cdot \overrightarrow{\mathbf{n}}$. Similarly, $d_2=-{\mathbf{b}}\cdot \overrightarrow{\mathbf{n}}$. $l_1$ and $l_2$ can be obtained by
    $l_i = |d_i-d|, i = 1,2$.

Following~\cite{BoltchevaRVD2017}, we approximate the 2D disk with a 10-vertex polygon and map it to the high-d space using Eq.~(4) in the main paper to define the tangential disk. Re-entrant clipping is then performed between the edges of this disk and the bisectors between site points in the target high-d space. Thus, RVCs are formed after the clipping. The set of $T$ of candidate triangles is then defined by the triangles which are dual to the restricted Voronoi vertices.

\section{Manifold Mesh Extraction}
As in~\cite{BoltchevaRVD2017,DEY2011483_cocone}, the result of the previous clipping step is a triangle soup. We apply a manifold extraction method following~\cite{BoltchevaRVD2017}. The candidate triangle set $T$ includes both the restricted Delaunay triangles $T_3$ and other supplementary triangles. The $T_3$ triangles are defined by triples of RVCs that are in mutual contact. 

The clipped RVC at a seed $\mathbf{\overline{x}}_i\in \mathcal{M}$, defined over $\mathcal{T}_{\mathbf{\overline{x}}_i}\mathcal{M}$ is constructed from RVC's vertices $\mathbf{\Lambda}_c$. Each Voronoi vertex $\mathbf{\Lambda}_c$ corresponds, in the dual sense, to a 2-simplex $\mathcal{K}_{i,j,k}$ with vertices $\{\mathbf{\overline{x}}_i, \mathbf{\overline{x}}_j, \mathbf{\overline{x}}_k\}$, where the Voronoi vertex is given by the intersection: $\mathbf{\Lambda}_c = \mathbf{D}^d_r(\mathbf{\overline{x}}_i)\cap \mathbf{B}(\mathbf{\overline{x}}_i,\mathbf{\overline{x}}_j)\cap \mathbf{B}(\mathbf{\overline{x}}_i,\mathbf{\overline{x}}_k)$, where $\mathbf{D}^d_r(\mathbf{\overline{x}}_i)$ is the high-d embedded intrinsic disk with the radius of $r$ at seed $\mathbf{\overline{x}}_i$, $\mathbf{B}$ is the bisector. The set $T_3$ consists of triangles that are consistently present in the local triangle sets of all three vertices: 
\begin{align}
    T_3 = \{\mathcal{K}_{i,j,k}| \mathcal{K}_{i,j,k}\in T(i)\cap T(j)\cap T(k)
    \}.
\end{align}
These triangles can be easily found, by generating all the index triples that correspond to the restricted Voronoi vertices, sorting and keeping the triples that appear three times in the sorted sequence. Note that in the sorted sequence, there are also triples that
appear once or twice. The corresponding set of triangles is referred to as $T_{12}$:
\begin{align}
    T_{12}=\{\mathcal{K}_{i,j,k}|\mathcal{K}_{i,j,k}\in (T(i)\cap T(j))\cup \nonumber \\
    (T(i)\cap T(k))\cup 
    (T(j)\cap T(k)) \setminus T_3 \}.
\end{align}
This means that a simplex $\mathcal{K}_{i,j,k}$ is included in $T_{12}$ if the RVC of $\mathbf{\overline{x}}_i$ intersects the bisectors associated with $\mathbf{\overline{x}}_j$ and $\mathbf{\overline{x}}_k$ but at least one of the RVCs of $\mathbf{\overline{x}}_j$ or $\mathbf{\overline{x}}_k$ does not reciprocate this relationship.

The manifold extraction procedure begins with the triangle set $T_3$ as the initial mesh. Triangles from $T_{12}$ are then considered for insertion sequentially. Each triangle is added only if its inclusion preserves the manifold property of the mesh; otherwise, it is discarded. Finally, we apply the post-processing step provided by Geogram~\cite{levy2015geogram} to fill any remaining holes from the previous step if necessary.

\section{Implementation Details}
Our neural high-d point embedding is implemented using PyTorch Geometric~\cite{Fey/Lenssen/2019}. Training is performed with the AdamW optimizer~\cite{loshchilov2018decoupled} using a base learning rate of 0.01 and a scaled learning rate of $10^{-4}$ for the attention modules. The network is trained with a batch size of 16 for 5000 epochs, employing a cosine learning rate scheduler with 30 warm-up steps.

The high-d particle sampling optimization is implemented in PyTorch~\cite{pytorch}. The optimization runs for 100 iterations, using a learning rate $10^{-3}$ and decayed by a factor of 0.6 every 20 iterations. The anisotropic manifold mesh reconstruction pipeline is implemented based on Geogram~\cite{levy2015geogram}. \textit{The source code and data will be publicly released after acceptance.}

\section{Data Augmentation and Testing Datasets}
\textbf{Data Augmentation.} To mitigate biases arising from geometric variations in the training set, especially in the direction and spatial distribution of local stretching ratios, we apply data augmentation through rotations (by $\pi/2$ around each of the three Euclidean axes), and reflections (with respect to the $xy$, $xz$, and $yz$ planes). These augmentations reorient the local stretching directions and redistribute the stretching ratios relative to the vertex coordinates. After applying this data augmentation strategy, the training set comprises 2,400 models of point clouds. This augmentation strategy follows the same rationale as in NASM~\cite{Li2024}, where such operations have been shown to enhance network performance and reduce directional bias in training data.

\textbf{Testing Datasets.} 
\textbf{(1) Baseline Sampling Dataset:} We first evaluate our method on a subset of the Thingi10k dataset~\cite{Thingi10K}, which comprises 80 diverse 3D shapes. The evaluation point clouds are directly randomly sampled from surface meshes, whose sampling procedure is the same as that of the training data. This experiment serves as a baseline for evaluating our method. The following three datasets are unseen datasets that are only used in the testing stage. \textit{All evaluations are conducted without any fine-tuning or retraining.} \textbf{(2) Synthetic Scanning Dataset:} We test on synthetic scanned point clouds that emulate real-world sensor data. We incorporate the Myles et al.'s dataset~\cite{Myles_2014}, which includes 3D shapes from the AIM@Shape and Stanford shape repository, offering rich diversity in geometric features and topological complexity. We select about 75 models from this dataset. The evaluation point clouds are generated through a scanning simulation pipeline using Blensor~\cite{blensor2011} as in~\cite{huang2024surface}. The scanned point clouds deviate slightly from the mesh-based surfaces and contain missing regions due to limited scanning angles, making this evaluation setting more challenging than the previous baseline dataset. \textbf{(3) Noisy and Irregular Sampling Dataset:} This task further evaluates our method's robustness and generalization by adding varying noise levels and irregular spatial sampling patterns into the above synthetic scanned point clouds to simulate degraded or non-uniform acquisition conditions. \textbf{(4) Large-Scale Dataset:} We also evaluate our method on point clouds with complex geometry details and scene-level datasets containing a significantly large number of points. For model-level point clouds, we select models from Myles et al.'s dataset~\cite{Myles_2014} and TetWeave dataset~\cite{binninger2025tetweave}, which contain intricate geometric structures; while for scene-level evaluation, we use ScanNet dataset~\cite{dai2017scannet}. ScanNet point clouds are derived from real scan data of indoor scenes.

\section{Additional Results and Evaluations}
\subsection{Our Results}
\begin{figure*}[t]
\centering
\includegraphics[width=0.98\textwidth]{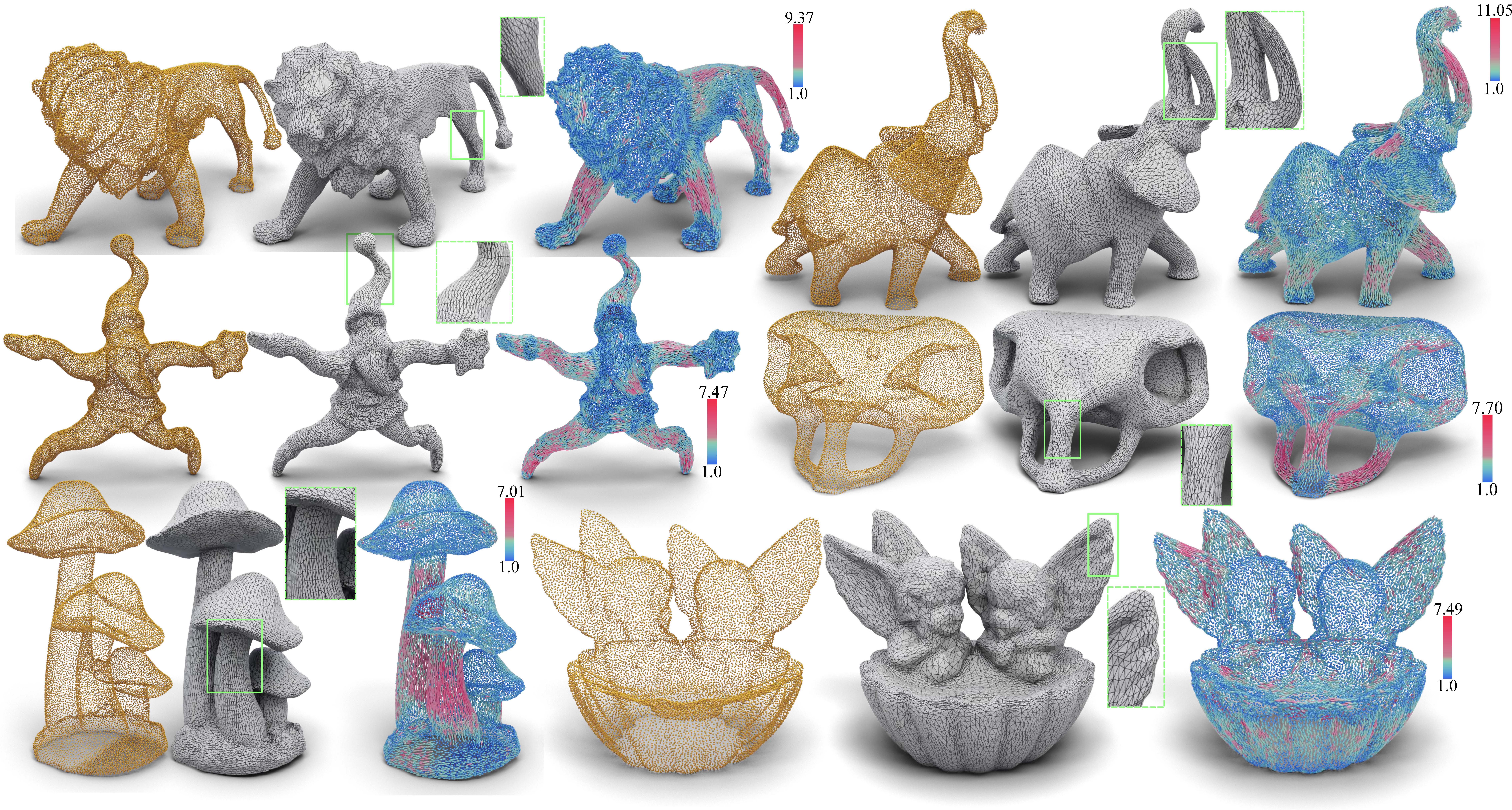}\vspace{0mm}
\caption{Additional our high-fidelity anisotropic surface mesh approximation results from input point clouds. The models are selected from Thingi10K dataset~\cite{Thingi10K} and unseen Myles et al.'s dataset~\cite{Myles_2014}. Left to right: input point clouds, reconstructed anisotropic meshes with zoom-in views, and estimated curvature tensors with corresponding stretching ratios for input point clouds denoted in colors.}\vspace{0mm}
\label{ours_add}
\end{figure*}

Fig.~\ref{ours_add} shows additional our HD-PEA method results from Thingi10K dataset and an unseen~\cite{Myles_2014} dataset. It is noted that our reconstructed high-fidelity anisotropic meshes can well capture geometric details and accurately approximate the surfaces from the input point clouds. 

The estimated curvature tensors from the neural high-d point embedding are visualized by ellipsoid-style glyphs, where the axes represent the principal curvature directions. The color encodes the corresponding anisotropy, defined by the stretching ratios of the tensors. \textit{To our knowledge, this is the first deep learning-based approach to compute the curvature tensor fields from the given unstructured point clouds.} 

We can see that the reconstructed anisotropic mesh elements well align with the estimated curvature tensors w.r.t. stretching ratios and directions. In the future, we will explore more downstream analyses and applications based on our HD-PEA  framework. 

\subsection{Comparison Results on Surface Reconstruction}
\begin{figure*}[t]
\centering
\includegraphics[width=0.98\textwidth]{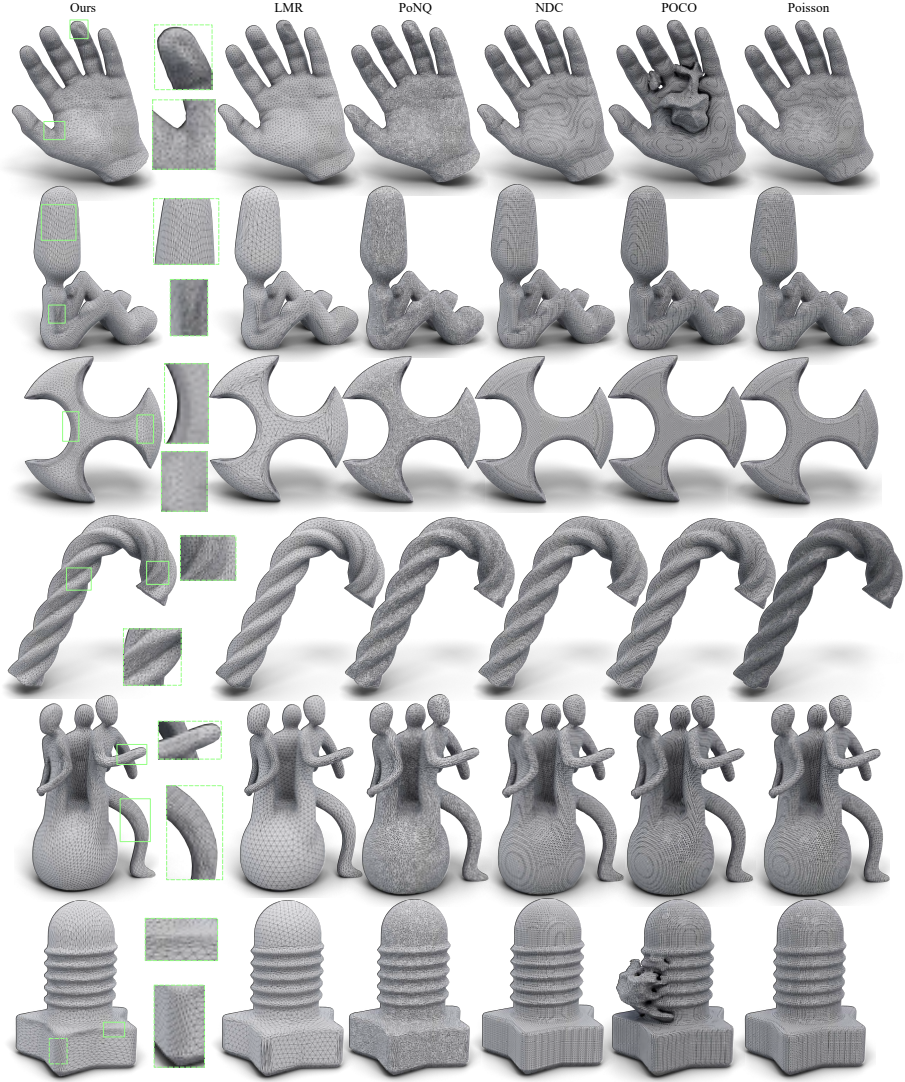}\vspace{0mm}
\caption{Additional comparison results of our HD-PEA method against state-of-the-art approaches. The point cloud models are selected from Thingi10K dataset (rows 2, 3, 4, and 6) and an unseen Myles et al.'s dataset~\cite{Myles_2014} (rows 1 and 5). Left to right: HD-PEA with zoom-in views, LMR, PoNQ, NDC, POCO, and Poisson. NDC, POCO, and Poisson use grid size of $128^3$.}
\label{mesh_compare_add}
\end{figure*}

Fig.~\ref{mesh_compare_add} shows additional comparison results with several state-of-the-art surface mesh reconstruction methods, including traditional and deep learning-based methods, such as Poisson~\cite{kazhdan2006poisson}, NDC~\cite{chen2022neural}, POCO~\cite{boulch2022poco}, PoNQ~\cite{maruani2024ponq}, and LMR~\cite{zhang2025high}. 

For Marching Cubes-like methods, such as Poisson, NDC, and POCO, often need very high grid resolutions to generate the final meshes in order to achieve a reasonable surface accuracy. Furthermore, they often exhibit zigzag artifacts on the surface due to their inflexible sampling directions. PoNQ employs QEM~\cite{garland1997surface} for surface approximation, it relies on Delaunay triangulation, which typically produces more vertices than desired as well as a relatively low mesh quality. LMR conforms to surface curvatures by using adaptive meshes; however, it does not adequately control stretching ratios for anisotropic triangles and relies on explicit curvature calculation, which is computationally expensive and vulnerates to generate a noticeable amount of degenerate triangles. 

As for our HD-PEA method (i.e., high-fidelity and lightweight anisotropic surface reconstruction), by stretching triangles along surface curvature directions with appropriate anisotropic ratios, it achieves a more accurate surface approximation that better aligns with the ground truth geometry with fewer mesh elements. Since our method does not require explicit computation of curvatures or directional fields, it offers enhanced efficiency and generalization compared to existing curvature-adaptive meshing approaches.

\subsection{Results on Large-Scale Complicated Geometry Objects and Scenes}
\begin{figure}[t]
\centering
\includegraphics[width=0.8\textwidth]{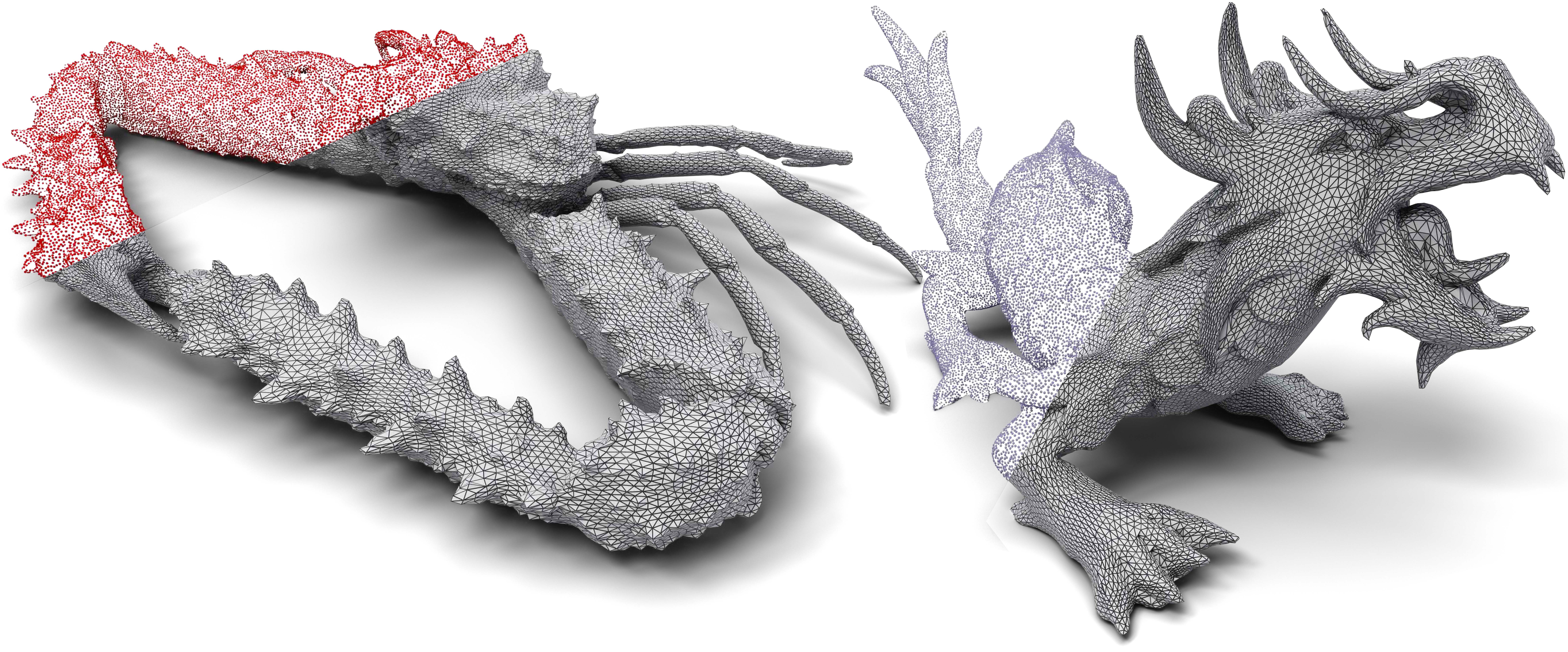}\vspace{-2mm}
\caption{Examples of reconstructed high-fidelity anisotropic meshes with preserving highly detailed geometric features (about 35K vertices) from the given large-scale point clouds (about 250K points) by our HD-PEA method.}\vspace{-8mm}
\label{teaser}
\end{figure}

\begin{figure*}[t]
\centering
\includegraphics[width=0.8\linewidth]{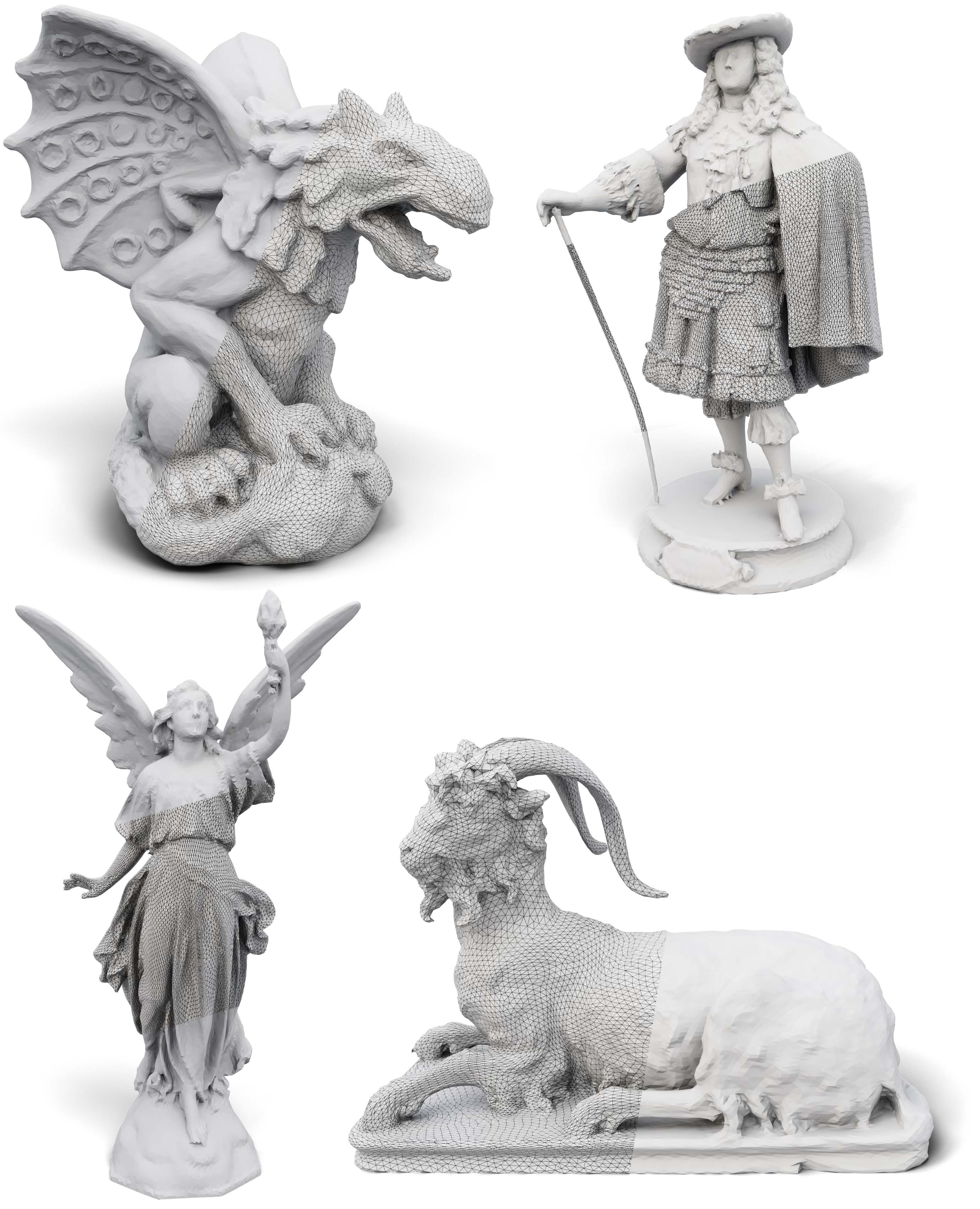}\vspace{-2mm}
\caption{Additional visualization of our large-scale reconstructed anisotropic mesh results (about 20K to 40K mesh vertices) on surface models (selected from unseen Myles et al.'s dataset~\cite{Myles_2014} and TetWeave dataset~\cite{binninger2025tetweave}) with complex geometry details.}\vspace{0mm}
\label{fig:complicated}
\end{figure*}

Fig.~\ref{teaser}, Fig.~\ref{fig:complicated}, and Fig.~\ref{fig:scannet} demonstrate that our method has strong performance on large-scale point clouds with complex geometry details on object-level and scene-level datasets, respectively, containing a significantly large number of points. These additional examples are selected from Myles et al.'s dataset~\cite{Myles_2014}, TetWeave dataset~\cite{binninger2025tetweave}, and ScanNet dataset~\cite{dai2017scannet}, even without fine-tuning or retraining. Fig.~\ref{scene_result} shows both qualitative and quantitative comparisons between our method and NDC on a ScanNet scene. Our method achieves a lower CD than NDC (0.786 vs. 0.814), while using only less than one-third of the mesh resolution. In this setting, we employ the proposed Patch-Based Meta-Embedding Inference (PMEI) to demonstrate the scalability and adaptability to large-scale data.

\begin{figure*}[t]
\centering
\includegraphics[width=0.98\linewidth]{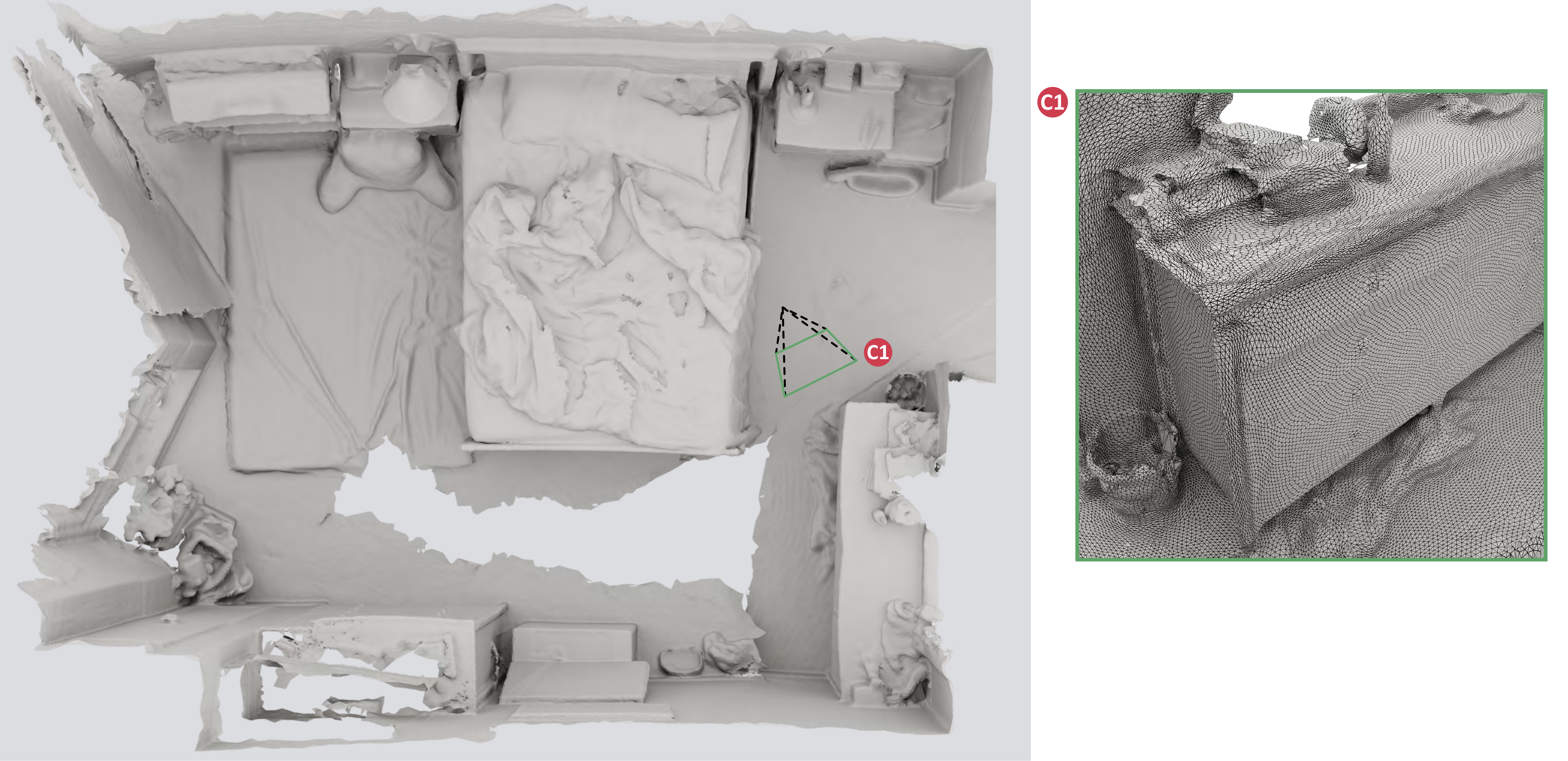}\vspace{0mm}
\caption{Additional visualization of our large-scale reconstructed anisotropic mesh result on another ScanNet scene (some holes exist due to the input scanned data). Some parts of walls and roofs are removed for visibility of interior geometry.
}\vspace{0mm}
\label{fig:scannet}
\end{figure*}

\begin{figure*}[t]
\centering
\includegraphics[width=0.98\linewidth]{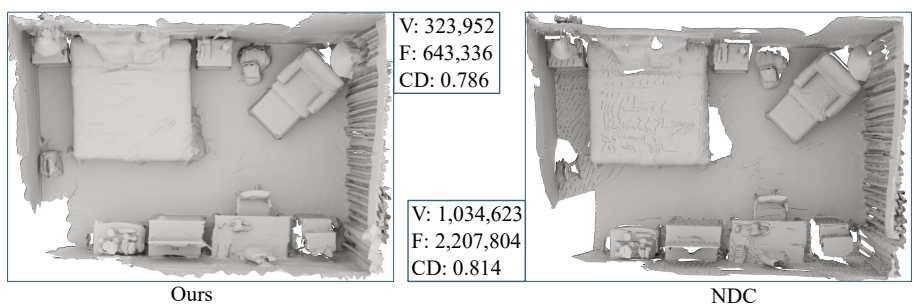}\vspace{0mm}
\caption{Qualitative and quantitative comparison of our approach against NDC, demonstrated on a scene model from ScanNet.}\vspace{-4mm}
\label{scene_result}
\end{figure*}

\subsection{Ablations on Input Normal and \# Points} 
Table~\ref{tab:normal_noise} shows that we added Gaussian noise with standard deviations of 0.1\% and 0.2\% to simulate two levels of normal-wise perturbations. Our method demonstrates good robustness to normal estimation errors. In the revision, we will add more ablation study on the normal orientation flips or inconsistency. Table~\ref{tab:comparison_points} shows the ablation study on reconstructed mesh quality and runtime under different numbers of input points. 
The performance remains stable and consistent across varying input sizes.
\begin{table}[h]
\caption{Quantitative evaluation of HD-PEA with normal noises on 75 synthetic scanned point clouds from Myles et al.'s dataset~\cite{Myles_2014}. The best results are highlighted in bold. Note: CD ($\times 10^{-5}$) and HD ($\times 10^{-2}$).
}\vspace{-4mm}
\label{tab:normal_noise}
\begin{center}
\resizebox{0.6\textwidth}{!}{
    \begin{tabular}{llllllllll}
    \toprule
   Method & $\#V_{out}$ & $\#f_{out}$ & CD $\downarrow$ & F1 $\uparrow$ & NC $\uparrow$ & HD $\downarrow$ & G $\uparrow$  \\
    \midrule
    Clean Normal & 7,028 & 13,989 & \textbf{0.649} & \textbf{0.989} & \textbf{0.979} & \textbf{0.798} & \textbf{0.702}\\
    \quad +0.1\% Noise & 8,761 & 17,513 & 0.673 & 0.988 & 0.978 & 0.828 & 0.701 \\
    \quad +0.2\% Noise & 9,517 & 19,029 & 0.674 & 0.988 & 0.972 & 0.868 & 0.701  \\
    \bottomrule
    \end{tabular}
}\vspace{-8mm}
\end{center}
\end{table}

\begin{table}[h]
\caption{Quantitative evaluation of HD-PEA with different numbers of input point samples on the dataset of Myles et al.}\vspace{-4mm}
\label{tab:comparison_points}
\begin{center}
\resizebox{0.6\textwidth}{!}{%
    \begin{tabular}{lllllllllll}
    \toprule
   Input Size & CD $\downarrow$ & F1 $\uparrow$ & NC $\uparrow$ & HD $\downarrow$ & G $\uparrow$ & Time (s) \\
    \midrule
    20,000 & 0.648 & 0.988 & 0.980 & 0.891 & 0.681 & 10.697\\
    40,000 (default) & 0.633 & 0.990 & 0.983 & 0.831 & 0.698 & 15.871  \\
    60,000 & 0.628 & 0.989 & 0.983 & 0.832 & 0.694 & 21.173 \\
    \bottomrule
    \end{tabular}%
}\vspace{-6mm}
\end{center}
\end{table}

\subsection{In-depth Analysis on Two-Stage NDC+NASM} 
Our HD-PEA reconstructs anisotropic surfaces directly from raw point clouds, while NASM performs anisotropic remeshing on an existing mesh. Unlike NASM, HD-PEA avoids any intermediate mesh generation stage. In our experiments (Fig. 7 in the paper), we use NDC($128^3$) to generate the input mesh for NASM, since NDC achieves the best performance among baseline reconstruction methods. However, the behavior of such two-stage pipelines (e.g., NDC+NASM) could be analyzed in greater depth, particularly regarding their sensitivity to intermediate mesh quality. As shown in Fig. 6 in the paper, NDC introduces severe aliasing artifacts that subsequently cause NASM to overfit these artifacts. If a lower-resolution reconstruction such as NDC($64^3$) is used instead, the intermediate mesh contains significantly larger reconstruction errors and aliasing artifacts, which would further propagate to the NASM remeshing stage and degrade the final results. In contrast, HD-PEA directly learns the anisotropic metric field in the high-dimensional embedding space from raw point clouds, avoiding this artifact-prone intermediate stage. This leads to more stable and faithful reconstructions, consistently achieving better surface accuracy (CD, F1, NC, HD) and anisotropic mesh quality metric G.

\subsection{Rendering and Roughness}
\begin{figure*}[t]
\centering
\includegraphics[width=0.95\textwidth]{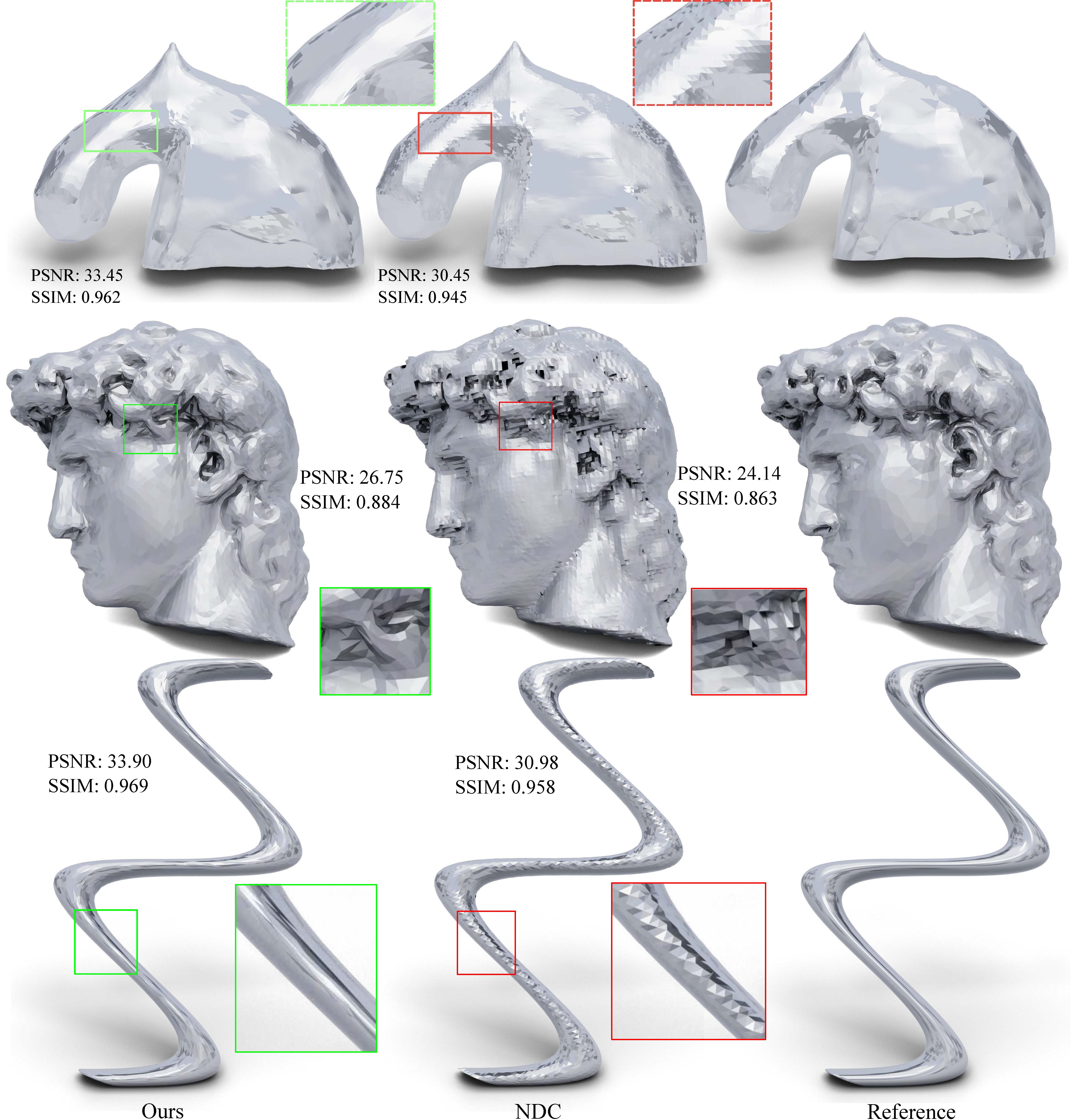}
\caption{Additional visual comparison of surface roughness. All renderings are shown under identical lighting condition. Note that the third row shows a sine-curved cylinder surface generated by an analytical function, providing a reference for the ideal roughness pattern expected under the same illumination. Compared with NDC~\cite{chen2022neural}, our HD-PEA method achieves more faithful approximations and produces a smoother surface in regions of local curvature with a much lower mesh resolution. Our results show higher PSNR and SSIM metrics than NDC results. The mesh vertex numbers are Moose model (HD-PEA: 5,193 vertices, NDC: 15,963 vertices, Reference: 4,068 vertices), David model (HD-PEA: 14,772 vertices, NDC: 27,774 vertices, Reference: 14,286 vertices) and Sine-Curved Cylinder model (HD-PEA: 4,992 vertices, NDC: 9,222 vertices, Reference: 16,450 vertices).}
\label{fig:roughness}
\end{figure*}

We further evaluate our HD-PEA method by analyzing its ability to preserve surface curvature characteristics, which directly influence perceived roughness or smoothness. It is the key factor for 3D shape designing and rendering. Recently, designers and architects have increasingly leveraged surface roughness for visual, aesthetic, and structural effects. In this context, roughness or smoothness arises from the local alignment of mesh edges and faces with respect to the underlying curvature of the surface. Fig.~\ref{fig:roughness} shows the visual comparison between our HD-PEA and NDC~\cite{chen2022neural} method (the second best result in our quantitative comparisons). However, NDC method is based on Marching Cubes, which introduces undersampling artifacts, such as zigzagging surface patterns, degrading surface smoothness due to the grid resolution constraint. This characteristic is especially favorable in the context of surface rendering and designing applications. Two image-based metrics, peak signal-to-noise ratio(PSNR) and structural similarity index measure (SSIM), are used for quantitative evaluation. Our results demonstrate much higher PSNR and SSIM metrics than NDC results in Fig.~\ref{fig:roughness}.  

\section{Illustration of Patch-Based Meta-Embedding Inference} 
\begin{figure*}[t]
\centering
\includegraphics[width=0.95\textwidth]{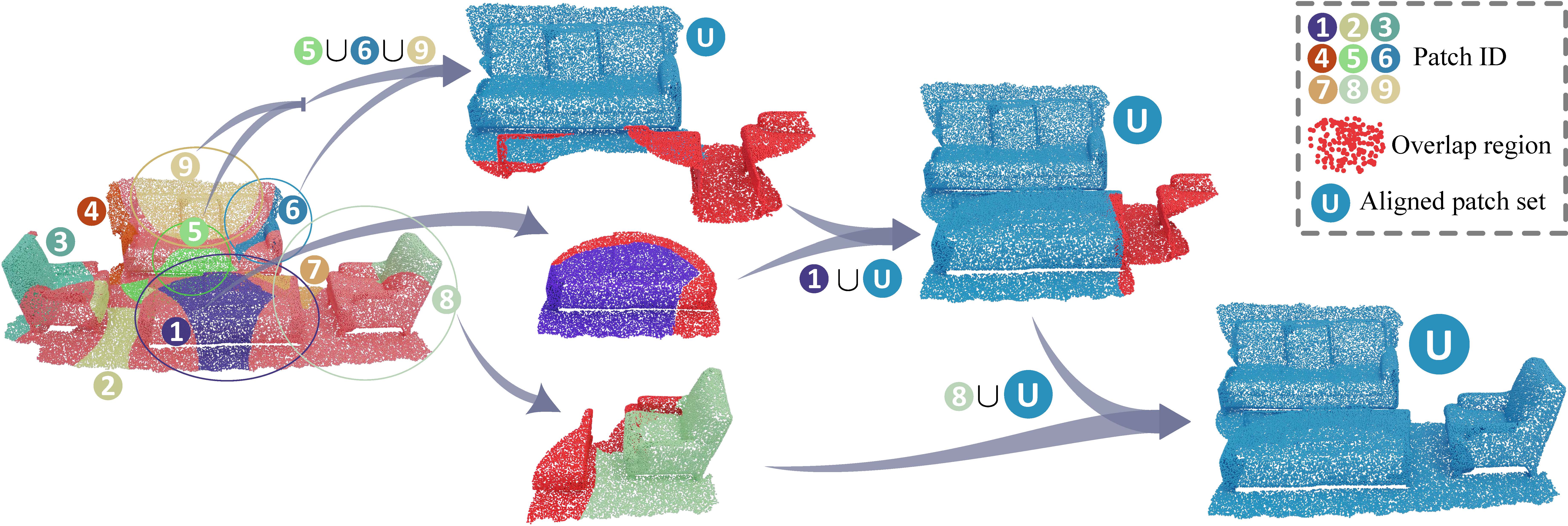}\vspace{0mm}
\caption{Illustration of the proposed patch-based meta-embedding inference (PMEI) on a large-scale scene-level point clouds.}\vspace{0mm}
\label{fig:psei} 
\end{figure*}

Fig.~\ref{fig:psei} demonstrates one example of our patch-based meta-embedding inference (PMEI) on a large-scale scene-level point clouds from ScanNet dataset~\cite{dai2017scannet}, i.e., an inference-time strategy that divides large point clouds into overlapping patches, processes each independently, and aligns their embeddings into a consistent global representation. It can efficiently handle arbitrarily large and complex point clouds without retraining while preserving embedding quality and consistency. The PMEI algorithm is provided in Algorithm~\ref{alg:mst_align}. The detailed technique is introduced in Section 3.1 of the main paper. 

\begin{algorithm}[t]
\footnotesize 
\setstretch{0.85}
\caption{Patch-Based Meta-Embedding Inference}
\label{alg:mst_align}
\begin{algorithmic}[1]
\Require Point cloud $\mathbf{P}\in \mathbb{R}^3$, 
number of patch centers $l$, patch size $k$, aligned neighbor count $t$.
\Ensure Globally aligned high-d embedding $\overline{\mathbf{P}}$

\State  $\mathcal{C} \gets \mathrm{FPS}(\mathbf{P},l)$, where $l<<N$. \Comment{patch centers}

\For{$j = 1 \to l$}
    \State $\mathbf{q}_j \gets \mathrm{KNN}_{\mathbf{p}}(\mathbf{c}_j, k)$      \Comment{construct patches}
    \State $\overline{\mathbf{q}}_j \gets \mathrm{PointTransformer}(\mathbf{q}_j)$ \Comment{compute high-d point embeddings}
\EndFor

\State $\mathrm{CenterAdjList} \gets$ empty list of length $l$
\For{$i = 1 \to l$} 
    \For{$j = 1 \to l, j \neq i$}
        \State $\mathrm{AdjList}[i].\mathrm{append}((j, \|\mathbf{c}_i - \mathbf{c}_j\|))$
            \Comment{store neighbor index and distance}
    \EndFor
    \State $\mathrm{AdjList}[i] \gets \mathrm{sort}(\mathrm{AdjList}[i], \text{key=distance})$
        \Comment{neighbors in ascending order of distance}
\EndFor

\For{$j = 1 \to n$}
    \State $\mathrm{processed}[j] \gets \text{false}$
\EndFor

\State Choose arbitrary root patch $\overline{\mathbf{q}}_r$
\State $\mathrm{AlignedSet} \gets \mathrm{AlignedSet} \cup \{\overline{\mathbf{q}}_r\}$
\State $\mathrm{processed}[r] \gets \text{true}$

\State Initialize empty min-priority-queue $\mathrm{PQ}$ \Comment{store (dist, (i,j))}

\For{$j = 1 \to n$}
    \If{$j \neq r$}
        \State $\mathrm{key} \gets \|\mathbf{c}_r - \mathbf{c}_j\|$
        \State $\mathrm{push}(\mathrm{PQ}, (\mathrm{key}, (r, j)))$
    \EndIf
\EndFor

\While{$\mathrm{PQ}$ not empty}
    \State $(\mathrm{dist}, (i, j)) \gets \mathrm{pop}(\mathrm{PQ})$
    \If{$\mathrm{processed}[j]$}
        \State \textbf{continue} \Comment{skip the already-aligned patches}
    \EndIf

    \For{$k=1 \to l$}
        \State $c_k \gets \mathrm{AdjList}[j][k]$
        \If{$\mathrm{processed}[\mathbf{c}_k] \text{ and } | \overline{\mathbf{Q}}_{\mathrm{aligned}}| < t$}
            \State $\overline{\mathbf{Q}}_{\mathrm{aligned}} \gets  \overline{\mathbf{Q}}_{\mathrm{aligned}} \cup \overline{\mathbf{q}}_{k} $
        \EndIf
    \EndFor

    \State $\mathcal{O}_{ij} \gets \mathbf{Q}_{\mathrm{aligned}} \cap \mathbf{q}_j$ \Comment{identify overlap in original point cloud}

    \State $\overline{\mathbf{q}}_i|_{\mathcal{O}_{ij}} \gets \overline{\mathbf{Q}}_{\mathrm{aligned}}(\mathcal{O}_{ij})$ 
    \Comment{select embeddings of overlapping points from aligned patch}
    \State $\overline{\mathbf{q}}_j|_{\mathcal{O}_{ij}} \gets \overline{\mathbf{q}}_j(\mathcal{O}_{ij})$
    \Comment{select embeddings of overlapping points from patch $j$}

    \State $(A[j], b[j]) \gets \arg\min\limits_{T_{ij}}\sum_{k=1}^m||\mathbf{p}_i^k-T(\mathbf{p}^k_j)||^2 $

    \State $\mathrm{AlignedSet} \gets \mathrm{AlignedSet} \cup \mathbf{A}[j]\cdot \overline{\mathbf{q}}_j + \mathbf{b}[j] $
    \State $\mathrm{processed}[j] \gets \text{true}$

    \For{$u = 1 \to n$}
        \If{not $\mathrm{processed}[u]$}
            \State $\mathrm{key} \gets \|\mathbf{c}_j - \mathbf{c}_u\|$
            \State $\mathrm{push}(\mathrm{PQ}, (\mathrm{key}, (j, u)))$
        \EndIf
    \EndFor
\EndWhile
    \State $\overline{\mathbf{P}} \gets \mathrm{scatter\_mean}(\mathrm{AlignedSet})$

\end{algorithmic}
\end{algorithm}

\end{document}